\documentclass[a4paper,11pt]{article}
\pdfoutput=1 

\usepackage{jheppub} 
\usepackage{amsmath,bm,amssymb}

\usepackage[T1]{fontenc} 

\title{\boldmath A critical point in the distribution of lepton energies from the decay of a spin-1 resonance}

\author[a,1]{L.~Bianchini\note{Corresponding author.}}
\author[a,b]{and G.~Rolandi}

\affiliation[a]{INFN, Sezione di Pisa, Pisa, Italy}
\affiliation[b]{Scuola Normale Superiore, Pisa, Italy}

\emailAdd{lorenzo.bianchini@pi.infn.it}
\emailAdd{Gigi.Rolandi@sns.it}

\abstract{
We consider a spin-$1$ resonance produced with an arbitrary spectrum of velocities and decaying into a pair of massless leptons, and we study the probability density function of the energy of the leptons in the laboratory frame.
A special case is represented by the production of $W$ bosons in proton-proton collisions, for which the energy of the charged lepton from the decaying $W$ can be measured with sufficient accuracy for a high-precision measurement of $M_W$.
We find that half of the resonance mass is a special value of the lepton energy, since the probability density function at this point is in general not analytic for a narrow-width resonance. In particular, the higher-order derivatives of the density function are likely to develop singularities, such as cusps or poles.
A finite width of the resonance restores the regularity, for example by smearing cusps and poles into local stationary points. The quest for such points offers a handle to estimate the resonance mass with much reduced dependence on the underlying production and decay dynamics of the resonance.}

\begin{document} 
\maketitle
\flushbottom

\section{Introduction}\label{sec:intro}

The problem of estimating the mass $M$ of a resonance that partially decays into undetectable particles often arises in collider experiments. For example, it occurs when some of the decay products of the resonance interact too weakly with the detector to produce a signal, or when they are measured with insufficient precision. If the kinematics of the collision event can be closed by other means, for example by using energy-momentum conservation, the problem has an obvious solution, otherwise it is under-constrained.

The loss of information due to the unobserved particles, which prevents $M$ from being unambiguously determined on an event-by-event basis, can be statistically recovered if the dymanics of both production and decay of the resonance are known.
When such a prior knowledge is available, the probability density function (p.d.f.) of the visible particle momenta $\{\bm p_\ell\}$, denoted by $\sigma^{-1}d\sigma/d\{\bm p_\ell\}$, can be computed by marginalizing the unobserved degrees of freedom. This marginal p.d.f. depends on the unknown resonance mass through the kinematics of the visible decay products. 
In general, the multi-dimensionality of the observable space makes the analytic calculation of this function of paramount complexity.  The problem is then best tackled by the use of Monte Carlo (MC) simulations of the process of interest, resulting in a discrete set of MC templates $\sigma^{-1}_{\rm MC}\,d\sigma^{\rm MC}(M_i)/d\{\bm p_\ell\}$ generated at different trial values of $M$. With these templates at hand, a numerical evaluation of the likelihood function of the data is possible, and the standard theory of likelihood-based estimators can be then used for estimating the unknown mass~\cite{James}.
By construction, such an approach is model-dependent, as it relies on theoretical assumptions (in fact, the complete $S$-matrix for the process of interest) for relating $\sigma^{-1}d\sigma/d\{\bm p_\ell\}$ to $M$. There are indeed examples where model uncertainties represent the limiting factor to the experimental accuracy. The determination of the $W$ boson mass at hadron colliders represents perhaps the most remarkable of such cases~\cite{WMassCDF, WMassD0, WMassATLAS}.

An alternative approach, which allows the aforementioned limitation to be partly overcome, consists in exploiting singularities in the phase-space of the visible observables~\cite{Kim}, i.e. special points where the tangent plane to the phase-space manifold is aligned with one of the invisible particle directions. 
The position of such pointed features in the spectra of kinematic variables can be related to the unknown mass, or, more generally, to combination of masses when there is more than one resonance in the decay chain~\cite{Han}. Besides being ideally independent from the details of the underlying dynamics, the main advantage of the phase-space singularity method is that a multi-dimensional problem is recast into a search for striking features, like sharp edges or cusps, on univariate distributions. A study of the phase-space singularity method in the context of the $W$ boson mass measurement at hadron colliders has been documented in Ref.~\cite{Rujula}. Not surprisingly, the optimal of such singularity variables is highly correlated with the usual transverse mass which, being a function of the transverse hadronic recoil, is affected by other well-known sources of experimental uncertainty~\cite{WMassATLAS}. 

Motivated by the need of reducing the model-dependence in the measurement of the $W$ boson mass without having to rely on the hadronic recoil, we will concentrate hereafter on the special case of a spin-1 resonance that decays into a pair of massless leptons, of which one is assumed to be measured with high precision, whereas the other is undetected. It has been pointed out in Ref.~\cite{Franceschini} that a two-body decay kinematics of this type features an obvious, yet subtle, invariance under boosts. Indeed, the mass of the mother particle plays a special role in the distribution of energy $E$ of the visible daughter particle. In particular, it can be proved that $M/2$ is a local maximum of the energy distribution $\sigma^{-1}d\sigma/dE$, if the mother particle is produced unpolarized. In this case, one can just measure $M$ by locating the point in the observed energy spectrum featuring the largest density. An application of this technique in the context of the top-quark mass measurement 
has been documented in Ref.~\cite{CMS-top}.

The argument leading to the identity $\mbox{argmax}[d\sigma/dE]=M/2$ relies on the assumption that the resonance is unpolarized. Instead, we would like to be as agnostic as possible with respect to the mechanism of production and decay of the resonance. In this spirit, we will study the mathematical properties of the p.d.f. of the lepton energy in full generality by deriving exact results in the approximation of a narrow-width resonance. Strictly speaking, any unstable resonance has a finite width $\Gamma>0$. In practice, the latter has to be compared with the experimental resolution $\sigma_E$ on the visible particle energy, which sets the minimum granularity at which differential properties of the p.d.f. $\sigma^{-1}d\sigma/dE$ can be defined. The case $\Gamma/\sigma_E\ll 1$, is then mathematically equivalent to treating the resonance in the narrow-width approximation. We will then validate the results against selected toy examples of production and decay dynamics. The results of this study motivate the usage of stationary points in the higher-order derivatives of the energy p.d.f., in particular of the second derivative, as an estimator of the resonance mass. Finally, we will study this method in the context of the $W$ boson mass measurement at the LHC using a MC simulation of the reaction $pp\to W^\pm X$, $W^\pm \to \mu^\pm\nu_\mu$ in proton-proton ($pp$) collisions at a center-of-mass energy $\sqrt{s}=13$ TeV. Quantitative estimates of the statistical and of the dominant theoretical uncertainty affecting the newly proposed method of measurement are also provided.

\section{Kinematics in the laboratory frame}\label{sec:kin}

Let $E$ ($E^*$) be the lepton energy in the laboratory (center-of-mass) frame, and $c^*\equiv\cos\theta^*$ the cosine of the polar angle in the center-of-mass frame with respect to the mother particle velocity $\bm\beta$ in the laboratory. We also define $E_0=M/2$ and introduce the dimensionless parameters $x=E/E_0$, $y=E^*/E_0$, and $z=E/E^*=x/y$. The lepton energy in the laboratory is related to $E^*$ and $c^*$ via a Lorentz transformation that depends only on the boost factor $\gamma=(1-\beta^2)^{-\frac12}$, with $\beta=|\bm \beta|$, namely:
\begin{align} \label{eq:E}
E & = \gamma E^* \left( 1+ \beta c^* \right).
\end{align}
The distribution of energies in the center-of-mass frame is assumed to be described by a Breit-Wigner function:
\begin{align} \label{eq:BW}
{h}(y) = \frac{1}{\pi} \frac{\Delta}{(y-1)^2 + \Delta^2},
\end{align}
where $\Delta=\Gamma/2M$. Since we are mostly concerned with narrow-width resonances, i.e. $\Delta\ll1,$ we can safely neglect the fact that the function $h$ in Eq.~\eqref{eq:BW} should be truncated at $y=0$ to prevent the center-of-mass energy from becoming negative. In fact, Eq.~\eqref{eq:BW} coincides with the more correct relativistic Breit-Wigner distribution~\cite{PDG} only when $y\approx1$ (although it is somehow simpler for the calculations to use the non-relativistic version of Eq.~\eqref{eq:BW}, the results presented here do not depend on this assumption). Finally, we remark that this p.d.f. converges to the Dirac delta function in the limit $\Delta\to 0$.

From Eq.~\eqref{eq:E}, the domain of $z$ is found to be:
\begin{align} \label{eq:range}
z \in  \left[ \gamma - \sqrt{\gamma^2-1}, \gamma + \sqrt{\gamma^2-1} \right]
\end{align}
where the relation $\gamma^2\beta^2=\gamma^2-1$ has been used.
If $\gamma\neq1$, Eq.~\eqref{eq:E} can be inverted yielding:
\begin{align} \label{eq:change}
c^* = \frac{1}{\sqrt{\gamma^2-1}}\left( \frac{E}{E^*}  - \gamma \right) \;\;\; \rightarrow \;\;\; dE = \sqrt{\gamma^2-1} E^* dc^*,
\end{align}
which implies a linear relation between $c^*$ and $E$ at a fixed value of $\gamma$ and $E^*$.

In the center-of-mass frame of a spin-1 resonance decaying to a pair of spin-$1/2$ particles, the cosine of the polar angle of the decaying lepton with respect to a given quantization axis is described by a p.d.f. of the form~\cite{Mirkes}:
\begin{align} \label{eq:cstar}
\frac{1}{\sigma}\frac{d\sigma}{d c^*} = \frac{3}{8} \left[\left( 1 + \frac{A_0(\gamma)}{2} \right) + {A_4(\gamma)}c^* + \left( 1 - \frac32 A_0(\gamma) \right) {c^*}^2 \right],
\end{align}
where the angular coefficients $A_{0,4}$ have been introduced as arbitrary dimensionless functions of the boost factor $\gamma$. The $A_0$ coefficient controls the fraction of longitudinal polarization ($f_0$) and satisfies the requirement $0\leq A_0\leq2$, whereas $A_4$ regulates the fractions of left ($f_L$) and right ($f_R$) transverse polarization. For a pure $V-A$ interaction, the angular coefficients are related to the polarization fractions $f_{0,L,R}$, relative to the direction of flight of the resonance, by the linear relations:
\begin{align} \label{eq:fLfRf0}
f_0 = \frac{A_0}{2}, \;\;\; f_L = \frac14\left( 2 - A_0 \pm A_4\right), \;\;\; f_R = \frac14\left( 2 - A_0 \mp A_4\right),
\end{align}
where the choice of sign depends on the lepton charge. Special cases of Eq.~\eqref{eq:fLfRf0} are the values $(A_0,A_4)=(0,\pm2)$, which corresponds to a purely left/right polarized resonance, and $(2/3,0)$, which corresponds to an unpolarized resonance. By combining Eqs.~\eqref{eq:change}-\eqref{eq:cstar}, we obtain the conditional p.d.f. of $E$:
\begin{align} \label{eq:diffE}
& \frac{1}{\sigma}\frac{d\sigma}{dE}(E \; | \; \gamma, E^*)  = \frac{1}{\sigma}\frac{d\sigma}{d c^*}\left| \frac{dc^*}{dE} \right| = \\ \nonumber
& \frac{3}{8 E^*}\frac{1}{\sqrt{\gamma^2-1}}\left[  \left( 1 + \frac{A_0}{2} \right) + {A_4}\left(  \frac{{E}/{E^*}  - \gamma }{\sqrt{\gamma^2-1}} \right) + \left( 1 - \frac32 A_0\right)\left( \frac{{E}/{E^*}  - \gamma }{\sqrt{\gamma^2-1}}  \right) ^2 \right],
\end{align}
where the explicit dependence of the angular coefficients on $\gamma$ has been omitted for simplicity.
Multiplying both sides of Eq.~\eqref{eq:diffE} by the constant $E_0$, we obtain:
\begin{align} \label{eq:diffx}
\frac{d\sigma}{dx}(x \; | \; \gamma, y) = & \frac{3}{8y}\left[ \frac{1+\frac{A_0}{2}}{\left( \gamma^2-1 \right)^{\frac12}} + \frac{A_4}{ \left(\gamma^2-1\right) }\left( \frac{x}{y} - \gamma \right) + \frac{1-\frac32A_0}{\left( \gamma^2-1 \right)^{\frac32}}\left( \frac{x}{y} - \gamma \right) ^2  \right] , \\ \nonumber
& \times  I\left(\gamma - \sqrt{\gamma^2-1} \leq \frac{x}{y} \leq \gamma + \sqrt{\gamma^2-1} \right),
\end{align}
where $I(\cdot)=1$ if the argument is true, $0$ otherwise. The p.d.f. of $x$ can be now obtained by marginalizing both $\gamma$ and $y$. We assume $\gamma \sim g(\gamma)$ independently of $y$, which is usually appropriate for a narrow-width resonance. Under this assumption, we can write:
\begin{align} \label{eq:diffxfull}
f(x) = & \int \frac{dy}{y} \, {h}(y) \, \int_{\frac{1}{2}\left( \frac{x}{y} + \frac{y}{x}\right)}^{+\infty}d\gamma \, g(\gamma)  \\ \nonumber 
& \times \; \frac{3}{8}\left[ \frac{1+\frac12A_0(\gamma)}{\left( \gamma^2-1 \right)^{\frac12}} + \frac{A_4(\gamma)}{ \left(\gamma^2-1\right) }\left( \frac{x}{y} - \gamma \right) + \frac{1-\frac32A_0(\gamma)}{\left( \gamma^2-1 \right)^{\frac32}}\left( \frac{x}{y} - \gamma \right) ^2  \right]
\end{align}
The p.d.f. $g$ is positive-definite and normalized to unity: $\int_1^{\infty}d\gamma \, g(\gamma)=1$.
We first consider the case that $g$ is an analytic function everywhere, in particular at $\gamma = 1$ (the alternative case will be discussed later). Under this assumption, it can be replaced by its Taylor series centered at $\gamma=1$:
 \begin{align} \label{eq:gTalyor1}
g(1+\kappa) = g^{(0)} + g^{(1)}\kappa + \hdots,
\end{align}
where $\kappa\equiv\gamma-1\geq0$. Likewise, we assume that $A_{0,4}(\gamma)$ are analytic at $\gamma=1$ such that:
 \begin{align} \label{eq:gTalyor2}
A_{0,4}(1+\kappa) = A_{0,4}^{(0)}  + A_{0,4}^{(1)}\kappa + \hdots.
\end{align}

We now move to study the behavior of $f$ when $x\approx1$.
To this purpose, we expand the right-hand side of Eq.~\eqref{eq:diffxfull} in terms of a small parameter $\epsilon$, such that $x=1+\epsilon$. In this limit, we have:
 \begin{align} \label{eq:limits}
\frac12\left( x + \frac1x \right) = 1 + \frac{\epsilon^2}{2} + {\cal O}(\epsilon^4), \;\;\; \left( \gamma^2 - 1 \right)^{-\frac{k}{2}} \approx {2}^{-\frac{k}{2}}\kappa^{-\frac{k}{2}}
\end{align}
where $k$ is an integer. 

\subsection{The narrow-width approximation}

We first consider the case of a narrow-width resonance, i.e. we set ${h}(y)=\delta(1-y)$. After integrating-out $y$, the right-hand side of Eq.~\eqref{eq:diffxfull} can be rewritten symbolically as:
\begin{align} \label{eq:diffxdelta}
f(1+\epsilon) = \int_{{\epsilon^2}/{2}}^{\delta}d\kappa \, \mbox{Pol}_2\left(\epsilon \, ; \, \kappa \, , g^{(k)} \, , A_{0,4}^{(k)}\right) + \int_{\delta}^{+\infty} d\kappa \, \mbox{Pol}_2\left(\epsilon \, ; \, \kappa \, , g(\kappa), \, A_{0,4}(\kappa) \right),
\end{align}
where $\mbox{Pol}_2(\epsilon \, ; \, \cdot)$ stands for a second-order polynomial in $\epsilon$. In Eq.~\eqref{eq:diffxdelta}, the integration region has been split into two disjoint intervals: $[\epsilon^2/2, \delta]$, where the cut-off $\delta$ is sufficiently small that the approximations in Eq.~\eqref{eq:gTalyor1}-\eqref{eq:gTalyor2} are valid to first order, and the complementary interval $[\delta,+\infty]$.
The first integral provides the contribution inside a neighborhood of $x=1$ from the phase-space region $\gamma\approx 1$, i.e. when the decaying particle is almost at rest; the second integral accounts for the contribution stemming from larger boost values. By virtue of the spin-1 assumption, the integrand function within both integrals is a quadratic polynomial in $\epsilon$, hence it has vanishing derivatives beyond the second order.
We can now compute explicitly the first integral at the right-hand side of Eq.~\eqref{eq:diffxdelta}. After a straightforward integration, we get:
\begin{subequations}\label{eq:diffxdelta1}
\begin{align}  \label{eq:diffxdelta1:1}
& \int_{{\epsilon^2}/{2}}^{\delta}d\kappa \, \left( g^{(0)} + g^{(1)}\kappa \right)  \left[ 1+\frac12\left(A_0^{(0)} + A_0^{(1)}\kappa  \right) \right] 2^{-\frac12}\kappa^{-\frac12} = \\ \nonumber
& = -g^{(0)}\left( 1 + \frac{A_0^{(0)}}{2} \right)|\epsilon| - \frac{1}{6} \left[ \frac{g^{(0)}}{2} A^{(1)}_0 + g^{(1)}\left( 1 + \frac{A_0^{(0)}}{2} \right) \right] |\epsilon|^3 +  K_\delta + \mathcal{O}(\epsilon^5) \\ 
\label{eq:diffxdelta1:2}
& \int_{{\epsilon^2}/{2}}^{\delta}d\kappa \, \left( g^{(0)} + g^{(1)}\kappa \right) \left(A_4^{(0)} + A_4^{(1)}\kappa \right) 2^{-1} \left(-1 + \epsilon \kappa^{-1}\right)  = \\\nonumber
& = \frac12\left[ {g^{(0)}} A_4^{(0)} \ln\delta + \delta\left( g^{(0)}A_4^{(1)} + g^{(1)}A_4^{(0)}\right) + \frac14g^{(1)}A_4^{(1)}\delta^2 \right] \epsilon  + \\ \nonumber
& -g^{(0)} {A_4^{(0)}}\epsilon\ln|\epsilon| + \frac{g^{(0)}}{4}A_4^{(0)}\epsilon^2 - \frac14\left( g^{(0)}A_4^{(1)} + g^{(1)}A_4^{(0)} \right)\epsilon^3 + K_\delta^\prime + \mathcal{O}(\epsilon^3) \\ 
\label{eq:diffxdelta1:3}
& \int_{{\epsilon^2}/{2}}^{\delta}d\kappa \, \left( g^{(0)} + g^{(1)}\kappa  \right) \left[ 1- \frac32\left( A_0^{(0)} + A_0^{(1)}\kappa + \hdots \right) \right]2^{-\frac32} \kappa^{-\frac32} \left( \kappa^2 - 2\epsilon\kappa + \epsilon^2\right) = \\  \nonumber
& = -\sqrt{2}g^{(0)}\left( 1- \frac32A_0^{(0)}\right) \delta^{\frac12}\epsilon + g^{(0)}\left( 1- \frac32A_0^{(0)}\right)|\epsilon|+ g^{(0)}\left( 1- \frac32A_0^{(0)}\right)|\epsilon|\epsilon \\ \nonumber
& \left[- \frac{\delta^{-\frac12}}{\sqrt{2}}g^{(0)}\left( 1- \frac32A_0^{(0)}\right) + \frac{1}{\sqrt{2}} \left( -\frac32 g^{(0)}A_0^{(1)} + g^{(1)}\left( 1- \frac32A_0^{(0)}\right) \right)\delta^{\frac12} \right] \epsilon^2 + \\ \nonumber
& - \frac12\left[ -\frac32 g^{(0)}A_0^{(1)} + g^{(1)}\left( 1- \frac32A_0^{(0)}\right) + \frac16g^{(0)}\left( 1- \frac32A_0^{(0)}\right) \right]|\epsilon|^3 + K_\delta^{\prime\prime} + \mathcal{O}(\epsilon^3)
\end{align}
\end{subequations}
where $K_\delta$ are constants that depend only on the cut-off $\delta$. By rearranging the various terms in Eq.~\eqref{eq:diffxdelta1}, we finally obtain:
\begin{align} \label{eq:diffxdelta2}
f(1+\epsilon)&  \approx A + B\epsilon + C|\epsilon| + D\epsilon^2 + E|\epsilon|\epsilon + F\epsilon^3 + G|\epsilon^3| + H\epsilon\ln|\epsilon| + \mathcal{O}(\epsilon^3),
\end{align}
where the constants $A,\hdots,H$ are independent of $\epsilon$. There are terms in this expansion which are not analytic at $\epsilon=0$. They are are proportional to the constants:
\begin{subequations}\label{eq:diffxdelta3}
\begin{align} 
\label{eq:diffxdelta3:1}
C & = -2g^{(0)}A_0^{(0)} \\
\label{eq:diffxdelta3:2}
E & = g^{(0)}\left( 1- \frac32A_0^{(0)}\right) \\
\label{eq:diffxdelta3:3}
G &= \frac23\left( g^{(0)}A_0^{(1)} + g^{(1)}A_0^{(0)}\right) -\frac{g^{(0)}}{12}\left(1 - \frac32A_0^{(0)}  \right)- \frac23g^{(1)}   \\
\label{eq:diffxdelta3:4}
H & = -{g^{(0)}}A_4^{(0)}.
\end{align}
\end{subequations}

As a consequence of Eq.~\eqref{eq:diffxdelta2}, the higher-order derivates of $f$ can develop various types of singularity at $x=1$: kinks or cusps (from $|\epsilon|$ terms), discontinuities (from $\mbox{sign}(\epsilon)$ terms), delta functions (from the derivatives of the latter), and poles (from the derivative of the $\epsilon\ln|\epsilon|$ term). This non-regular behavior should not come as a surprise: even if $g$, $A_0$, and $A_4$ are smooth functions, the transformation in Eq.~\eqref{eq:change} becomes singular in the limit $\gamma\to1^+$. When convoluted with a continuous spectrum of boosts, this primordial singularity is weighted by an infinitesimal cross section $g(1)dx$, but still percolates to the final p.d.f., in a way that depends on how the phase-space $(\gamma, \, c^*)$ gets populated.
We anticipate that the appearance of a singularity in a strict mathematical sense is a consequence of treating the resonance in the narrow-width approximation. Within this approximation, however, its existence is a robust result, as discussed later.

The nature of such singularity is, to some extent, akin to the phase-space singularity discussed in Ref.~\cite{Kim}. Indeed, for a fixed value $\gamma>1$, the variable $x$ has two wall singularities associated with the decay of the visible particle collinear or anti-collinear with the velocity of the resonance. These configurations correspond to edges of the phase-space. When $\gamma= 1$, a singularity of higher degree appears because the dimensionality of the phase-space shrinks from a line to a point. The singularity studied here has, however, a richer phenomenology compared to the algebraic singularity of Ref.~\cite{Kim}, because it depends not just on the geometry of the phase-space manifold, but also on how the dynamics of production and decay distributes events across the phase-space.

It is interesting to consider some limiting cases of Eq.~\eqref{eq:diffxdelta3}. As expected, an unpolarized resonance gives rise to a p.d.f. of the form $f(x) = A + C|x-1|+\mathcal{O}((x-1)^2)$, implying that $x=1$ is a local maximum of the density (in particular, it is a cusp if $g^{(0)}>0$ and a stationary point otherwise). This is in agreement with the result obtained in Ref.~\cite{Franceschini}.
Whenever the boost p.d.f. has a vanishing amplitude in the neighborhood of $\gamma=1$,  i.e. $g^{(k)}=0$ for the first $k$ derivatives, the coefficients in Eq.~\eqref{eq:diffxdelta3} are also vanishing. A special case is when there is a minimum momentum threshold on the production of the resonance, such that $g(\gamma)\equiv0$ for $\gamma\leq\gamma_{\rm thr.}$ (in this case, all the derivatives at $x=1$ are formally zero). The expansion of Eq.~\eqref{eq:diffxdelta2} thus contains only terms of order $\epsilon^k$, with $k=0,1,2$: in the neighborhood of $x=1$, the energy p.d.f. is proportional to a parabola. Equation~\eqref{eq:range} can be then used to express $E_0$ in terms of the lower ($E_-$) and upper ($E_+$) edges of the interval in which $f(x)\propto \mbox{Pol}_2(x)$, as:
\begin{align} \label{eq:threshold}
E_0 = \sqrt{E_-E_+}.
\end{align}

We now briefly consider the possibility that either of the functions in the integrand is not analytic at $\gamma=1$, such that the Taylor expansion of Eq.~\eqref{eq:gTalyor1}-\eqref{eq:gTalyor2} are not defined. To fix the ideas, we consider the case $g(\gamma)\sim (\gamma-1)^\alpha$, with $\alpha>0$, for which $\gamma=1$ is a cusp point. As we will see later, this choice of p.d.f. finds at least one remarkable physical application. In this case, Eq.~\eqref{eq:diffxdelta1} gets modified by the appearance of terms like $|\epsilon|^{2\alpha+m}$, where $m=1,2,3,\hdots$ is an integer, which, for arbitrary values of $\alpha$, gives rise to the same phenomenology of non-regularity on $f$.

\subsection{Finite-width effects}
The regularity of laboratory energy p.d.f. is restored by integrating over a continuous spectrum of center-of-mass energies. Indeed, in the limit $\gamma\to1^+$, the laboratory frame coincides with the center-of-mass frame, i.e. $f(x)=h(x)$, which is a smooth function. In the case of a finite width, Eq.~\eqref{eq:diffxdelta} applies with the replacement:
\begin{align} \label{eq:xtoz}
\epsilon\to \epsilon^\prime = z-1 = \frac{x}{y}-1 \equiv \frac{1+\epsilon}{y} -1,
\end{align}
where $\epsilon$ is again defined as $\epsilon=x-1$. Equation~\eqref{eq:diffxdelta} thus becomes:
\begin{align} \label{eq:diffxdelta_average}
f(1+\epsilon) \approx  \int \frac{dy}{y} {h}(y) \, f\left(\frac{1+\epsilon}{y}\right).
\end{align}
Consider for example a term like $|\epsilon^\prime|$ in the expansion of $f$. Upon integration over $y$, its first derivative calculated at $x=1$ gives:
 \begin{align} \label{eq:smooth}
 \frac{d}{d\epsilon} \left[ \int \frac{dy}{y} {h}(y) \left| \frac{1+\epsilon}{y} -1  \right|\right]_{\epsilon=0} = \int \frac{dy}{y^2} {h}(y) \, \mbox{sign}(1-y)  = \mathcal{O}(\Delta^2),
\end{align}
hence the new minimum/maximum of $f$ gets displaced from $x=1$ by an amount of $\mathcal{O}(\Delta^2)$. Notice that if the $k$-th order derivative has a kink such that $|\lim_{x\to 1^+}f^{(k)}| \neq |\lim_{x\to 1^-}f^{(k)}|$, the integration over $y$ smears this singularity into a stationary point whose position depends not just on $\Delta$, but also on $g$ and $A_{0,4}$, which determine the left and right slopes of $f^{(k)}$. In the latter case, nothing can be said {\it a priori} about $E_0$, unless that it must be close to the stationary point.
For example, in the case of a symmetric kink in the second-order derivative, such a displacement is of $\mathcal{O}(\Gamma^2/M)$ and could be in principle subtracted from the measured stationary point, if $\Gamma$ were also known. In general, by knowing the width of the resonance and by relying on some theory prior on the coefficients of Eq.~\eqref{eq:diffxdelta2}, the measured stationary point can be calibrated to recover an unbiased estimator of $E_0$.

\subsection{An explicit example}\label{sec:WLHC}

We now discuss the case of $W$ boson production at the LHC which allows us to specialize some of the generic formulas derived before.

By using the fact that $d\gamma^2=d(|\bm q|^2/M^2)$, we can write
\begin{align} \label{eq:WatLHC}
\lim_{\gamma\to 1^+} g(\gamma) & \propto  \lim_{\gamma\to 1^+}  \frac{d\sigma}{d|\bm q|^2} = \lim_{\gamma\to 1^+}  \int  dq^2_{\rm T} \, dq^2_z \frac{d^2\sigma}{dq^2_{\rm T} \, dq^2_z}\delta(|\bm q|^2 - q^2_{\rm T} - q^2_z) \nonumber \\
& =  \lim_{\gamma\to 1^+}  \int_0^{|\bm q|^2} dq^2_{\rm T} \frac{d^2\sigma}{dq^2_{\rm T} \, dq^2_z}=  \lim_{\gamma\to 1^+}  \int_0^{|\bm q|^2} dq^2_{\rm T} \frac{1}{2E\sqrt{ |\bm q|^2 - q^2_{\rm T} }} \frac{d^2\sigma}{dq^2_{\rm T} dy} \\ \nonumber
& \approx \left[\frac{d^2\sigma}{dq^2_{\rm T} dy}\right]_{0}  \frac{|\bm q|}{M} \int_0^1 d\zeta \frac{1}{\sqrt{1-\zeta}} \propto  \sqrt{\gamma^2-1},
\end{align}
with $\zeta=q_{\rm T}/|\bm q|$. Hence, $g(1)=0$. Furthermore, since $g\approx \sqrt{\gamma^2-1}$, the boost spectrum is not analytic at $\gamma=1$.
The finiteness of the double-differential cross section $\left[d^2\sigma/dq^2_{\rm T}dy\right]_0$, where $y$ is the rapidity of the $W$ boson, is a general result that arises from the small transverse momentum behavior of cross sections in hard processes~\cite{Petronzio}:
\begin{align} \label{eq:sudakov}
\frac{d\sigma}{dq^2_{\rm T} }(q_{\rm T} \, | \, y) = \sigma_0 + \sigma_1 q_{\rm T}^2 + \hdots
\end{align}
Similarly, the differential cross section $d\sigma/dy$ at $y=0$ is finite because it is proportional to the product of the partonic densities evaluated at $x_{1,2}=M/\sqrt{s}$, where $\sqrt{s}$ is the proton-proton center-of-mass energy. By using a complete Monte Carlo simulation of this reaction, discussed in Sec.~\ref{sec:W}, we also find that the $\sim(\gamma-1)^{\frac12}$ rise is limited to a tiny region of phase-space, typically $\gamma-1\lesssim5\times10^{-4}$, corresponding to the region $|\bm q|\lesssim 4$ GeV, i.e. where the differential cross section in $q_{\rm T}$ is rapidly growing. For $\gamma$ values in excess of about $10^{-3}$, the boost p.d.f. is well approximated by a power law of the form $g(\gamma)\sim (\gamma-p_0)^{-p_1}$, with $p_0\approx0.9$ and $p_1\approx 0.8$. 

The angular coefficients are dimensionless functions of $|\bm q|/M$ and $y$ encoding the average polarization of $W$ bosons produced in hadron collisions as a function of the $W$ boson kinematics~\cite{Mirkes}. They are simultaneously determined by the partonic density functions (PDFs) of the proton and by emission of additional QCD radiation. Furthermore, they depend on the reference frame, i.e. they are not rotation-invariant. The coefficients $A_{0,4}$ of Eq.~\eqref{eq:cstar} are calculated in a particular rest frame of the $W$ boson defined by a boost along the velocity $\bm\beta$, which we will refer to as the helicity frame. For fixed and small values of $\gamma$, their values are determined by the average of $A_{0,4}$ over all momenta $\bm q$ defining the surface of a sphere of radius $|\bm q|$, such that $\gamma \approx 1+ \frac{|\bm q|^2}{2M^2}$:
\begin{equation} \label{eq:WatLHC2}
\lim_{\gamma\to1^+} A_{0,4}(\gamma) = \lim_{|\bm q|\to 0} \frac{\int d^3\bm q \frac{d^3\sigma}{d^3\bm q} \delta( \gamma - \frac{|\bm q|^2}{2M^2} - 1) A_{0,4}(\bm q) }{\int d^3\bm q \frac{d^3\sigma}{d^3\bm q} \delta( \gamma - \frac{|\bm q|^2}{2M^2} - 1)}
\end{equation}
Similarly to Eq.~\eqref{eq:WatLHC}, one can easily verify that all directions $\bm q/|\bm q|$ are equally likely in the limit $|\bm q|\to 0$:
\begin{equation} \label{eq:WatLHC3}
\lim_{|\bm q|\to 0} \frac{d\sigma}{d|\bm q|d\Omega} = \frac{|\bm q|^2}{2M_W} \left[\frac{d^3\sigma}{dq^2_{\rm T}\, dy\, d\phi}\right]_{0}, 
\end{equation}
where the right-hand side does not depend on the direction $\Omega$. When $|\bm q|$ is small, the direction of the quark and antiquark in any rest frame of the $W$ boson remain almost anti-parallel. They both carry spin parallel or anti-parallel to their respective momentum. While averaging over the full solid angle, their directions, which are fixed in the laboratory frame, move isotropically in the helicity frame around the quantization axis defined by $\bm q$ itself. The net result must be a uniform distribution in $c^*$, i.e. $A^{(0)}_{0}=2/3$ and $A^{(0)}_4=0$, or, using Eq.~\eqref{eq:fLfRf0}, $f_L=f_R=f_0$. This is only true when $|\bm q|\to 0$. For small finite momenta, a tiny polarization is produced. The amount by which a polarization is built by the misalignment of the quark directions can be estimated looking at a particular configuration where $y=0$ and $|\bm q|=q_{\rm T}$. In this case, the center-of-mass frame is related to the Collins-Soper (CS) frame~\cite{CS} by a rotation of $\pi/2$. From a MC simulation, we find that the longitudinal polarization in the CS frame is built at a pace of about $A_0(q_{\rm T})\approx 2\times(q_{\rm T}/M_W)^2$, or equivalently $A_0(\gamma)\approx 4\gamma$. The values of $(f_L,f_R, f_0)$ in the helicity frame is thus perturbed by an amount of similar size from the values $(\frac14,\frac14,\frac12)$ in the limit $|\bm q|\to 0$.
When $|\bm q|$ grows, large values of $q_{\rm T}$ in Eq.~\eqref{eq:WatLHC2} become increasingly unlikely and the cross section favors a longitudinal motion with $|q_z|\gg q_{\rm T}$. In this latter case, a net transverse polarization is built as a consequence of the PDF ratio $q/\bar q$  growing fast at large rapidities. Again using a MC simulation, we find an empirical slope $dA_4/dy\approx \pm0.3$ for $q_{\rm T}=0$, where the sign depends on the charge of the $W$ boson. Given that $dy=d\ln\gamma$, we also have $A_4^{(1)}\approx0.3$.

We notice that the same argument applies to the case of a proton-antiproton collider, for which $W$ bosons produced almost at rest in the laboratory frame are preferentially polarized in the direction of the antiproton. When integrating over the full solid angle, however, the average polarization in the helicity frame vanishes.  


\subsection{Discussion}

We now summarize the results obtained so far. When considering the two-body decay of a spin-1 resonance, the probability density function $f$ describing the laboratory energy of any of the two daughter particles, assumed to be massless spin-$\frac12$ particles, is in general not analytic at $x=1$, when the narrow-width approximation for the resonance is made. In particular, the derivatives of $f$ are likely to be non-derivable, discontinuous, or divergent at that point, depending on the boost factor p.d.f. and on the polarization of the resonance. The appearance of a local maximum of the density at $x=1$ is in general a fortuitous occurrence. A pole at $x=1$ in the first derivative is associated with the presence  of a non-zero transverse polarization at rest. Conversely, cusps in the second derivative appear quite naturally as a result of terms of the form $|x-1|^3$ in the expansion of $f$ around $x=1$, which do not compete with $(x-1)^3$ terms from higher boost factors. This is a general result that only depends on the spin-1 assumption for the resonance. The condition for which a cusp is generated amounts to $|G|>|F|$, as defined in Eq.~\eqref{eq:diffxdelta2}. If $g^{(0)}=0$, this is satisfied if $A_4^{(0)}=0$ (otherwise the coefficient $E$ in Eq.~\eqref{eq:diffxdelta3} would be non-zero, giving rise to a pole in $f^{(1)}$). If instead $g^{(0)}>0$ and $A_4^{(0)}=0$, the condition for developing a cusp in $f^{(2)}$ is:
 \begin{align} \label{eq:condition}
\left|A_4^{(1)}\right| < \frac83\left|\left( A_0^{(1)} + \frac{g^{(1)}}{g^{(0)}}A_0^{(0)}\right) -\frac{1}{8}\left(1 - \frac32A_0^{(0)}  \right)- \frac{g^{(1)}}{g^{(0)}}  \right|
\end{align}
which is satisfied if $A_4$ is a slowly varying function of $\gamma$. We remark that even if Eq.~\eqref{eq:condition} were not satisfied, kinks at $x=1$ for at least one among $f^{(0)}$, $f^{(1)}$, and $f^{(2)}$ will be present, so that a divergence in the higher-order derivatives will eventually show up.

The following search algorithm is then proposed. For simplicity, we assume the laboratory energy $E$ to be normalized to a constant $E_{0}^\prime$, playing the role of a trial mass. We define $x=E/E_{0}^\prime$ and set $f^{(0)}\equiv f$ for consistency of notation. Then:
\begin{enumerate}
\item if the resonance is known to be unpolarized, then define $\hat{x}_1 = \mathrm{argmax}[f^{(0)}]$ and stop, else compute $f^{(1)}$;
\item if a pole or cusp in $f^{(1)}$ is found, then define such point $\hat{x}_2$ and stop, else compute $f^{(2)}$;
\item if $f^{(2)}=\mathrm{const.}$ over a range $[x_-,x_+]$, then define $\hat{x}_3=\sqrt{x_-x_+}$ and stop. Else: if there is a cusp, define such point $\hat{x}_3$ and stop;
\item if no such points exist, then compute $f^{(k)}$, with $k\geq3$, and continue searching for a singularity $\hat{x}_{k+1}$ .
\end{enumerate}
The mass estimator is then defined as $\hat{M} = 2\hat{x}_kE_{0}^\prime$.
When a broad distribution of energies in the center-of-mass frame is accounted for, the analyticity of $f$ is restored. In particular, poles and cusps are regularized into local stationary points. These points are in general displaced from $x=1$ by an amount that vanishes in the limit $\Delta\to0$.
Furthermore, since there may be a multiplicity of such stationary points, a prior on $M$ will be in general needed to disambiguate among them and for an ultimate calibration of the estimator.
The determination of the unknown resonance mass is then recast as a univariate optimization problem, in a way that decouples from the details of the underlying production and decay dynamics to the extent that the resonance width can be neglected.

\section{Numerical examples}\label{sec:toy}

\begin{figure}[hbtp]
 \begin{center}
   \includegraphics[width=1\textwidth]{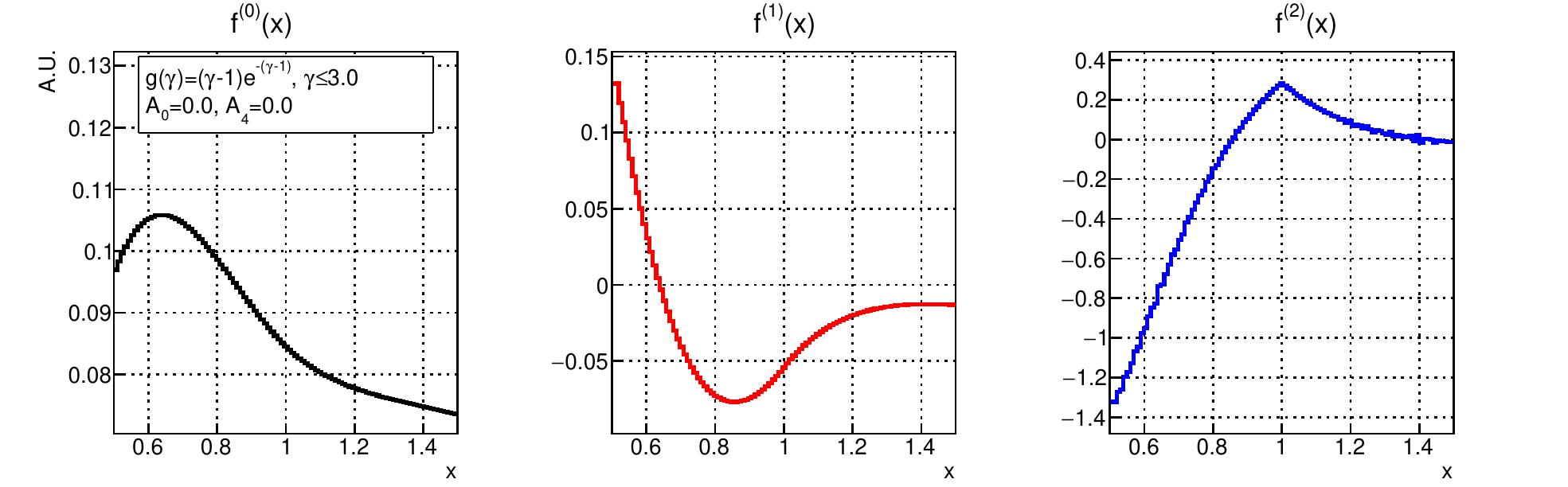}\\
   \includegraphics[width=1\textwidth]{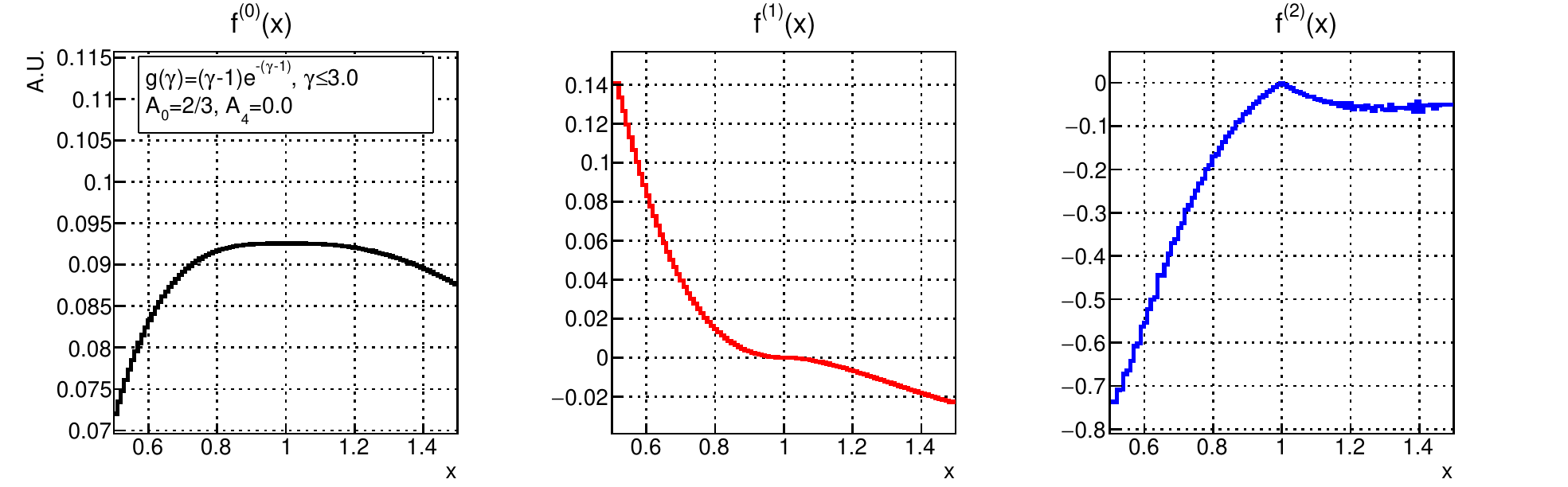}\\
   \includegraphics[width=1\textwidth]{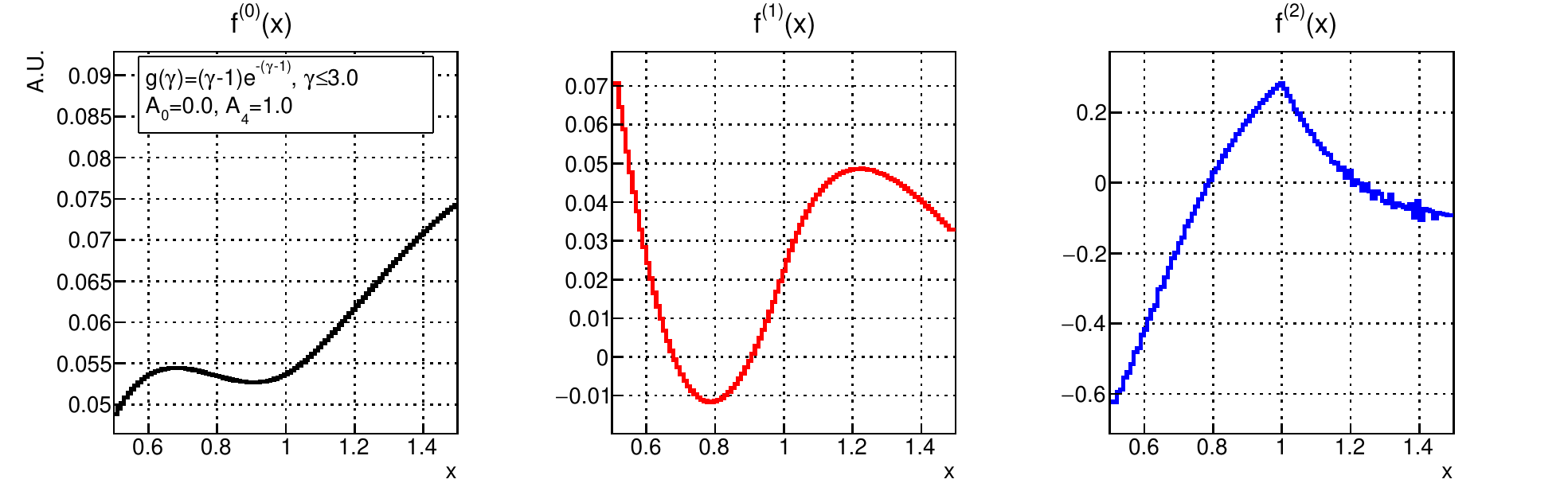}\\
   \includegraphics[width=1\textwidth]{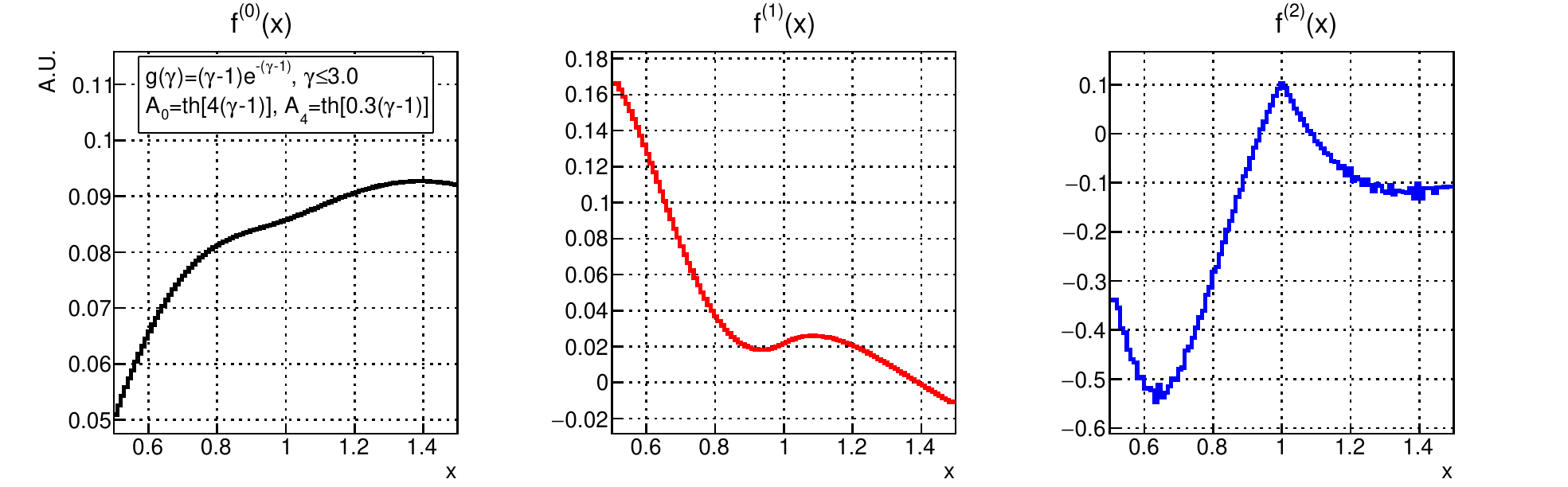}\\
   \caption{The lepton energy p.d.f $f\equiv f^{(0)}$ (left) of Eq.~\eqref{eq:diffxfull} with its first (center) and second (right) derivative for a boost p.d.f. of the form $g(\gamma)\propto(\gamma-1)\exp(-\gamma)$ and the choice $(A_0,A_4)=$ $(0,0)$ (first row), $(2/3,0)$ (second row), $(0,1)$ (third row), and $(\mbox{tanh}\left[4(\gamma-1)\right], \mbox{tanh}\left[0.3(\gamma-1)\right])$ (fourth row). The decaying resonance is treated in the narrow-width approximation.}
   \label{fig:toy1}
 \end{center}
\end{figure}

\begin{figure}[hbtp]
 \begin{center}
   \includegraphics[width=1\textwidth]{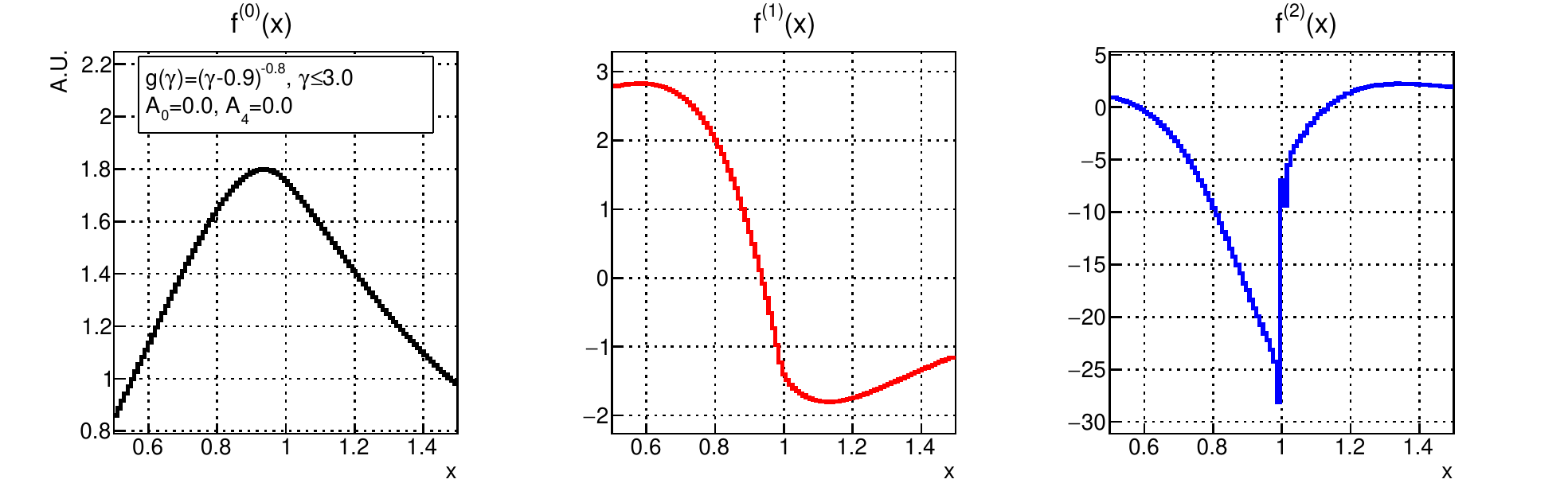}\\
   \includegraphics[width=1\textwidth]{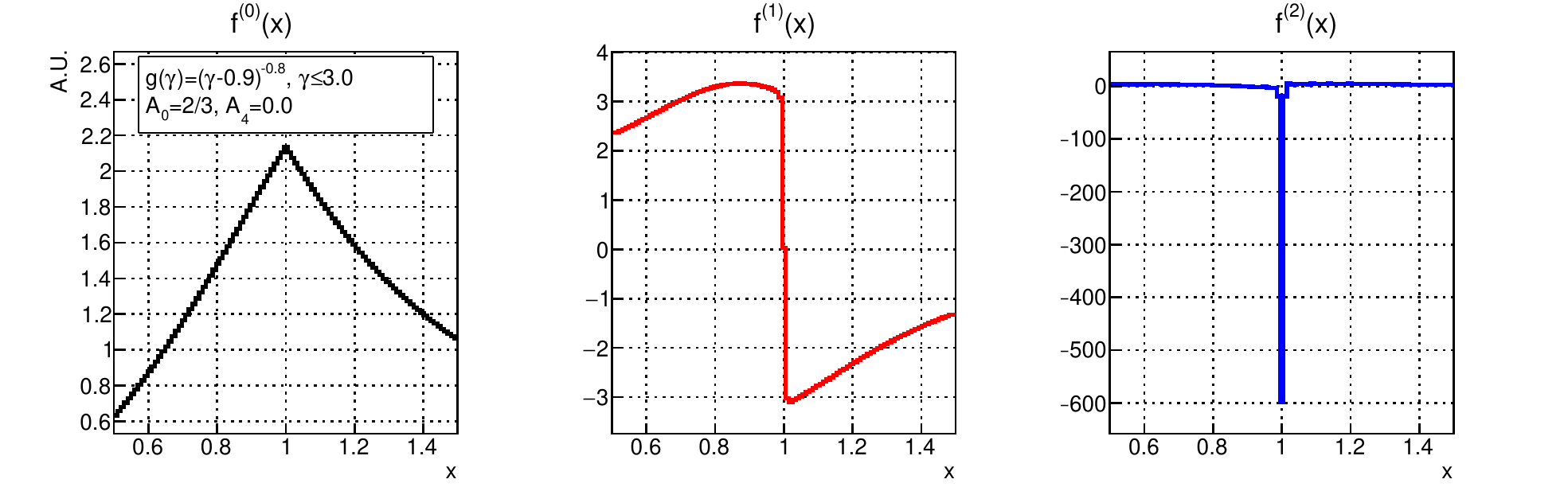}\\
   \includegraphics[width=1\textwidth]{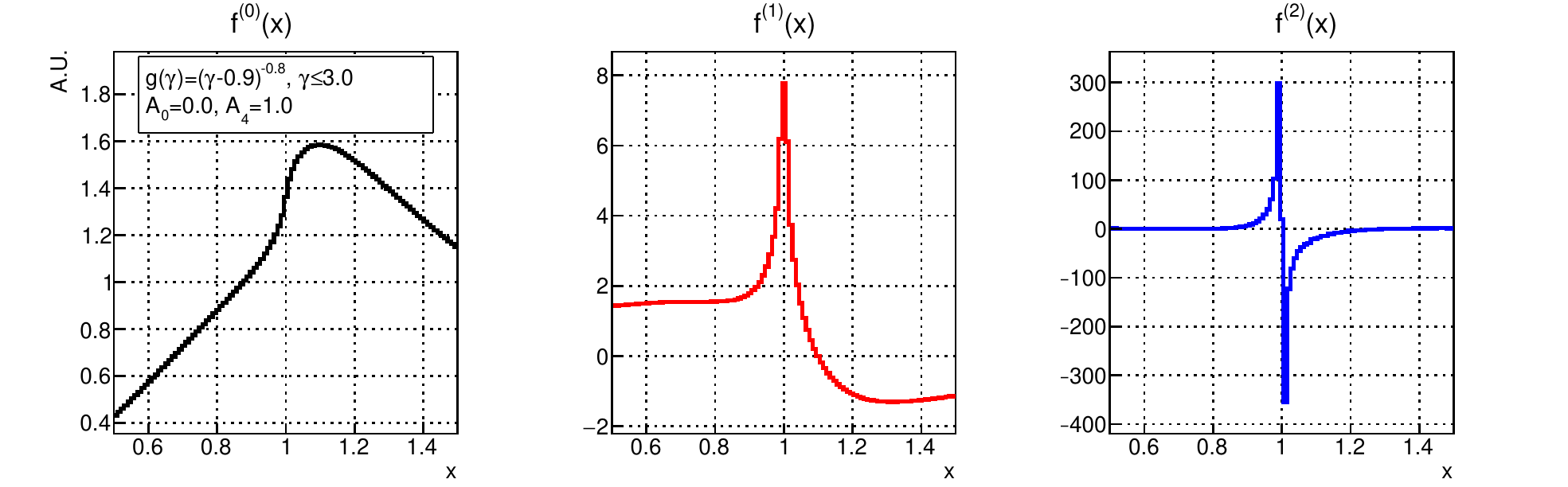}\\
   \includegraphics[width=1\textwidth]{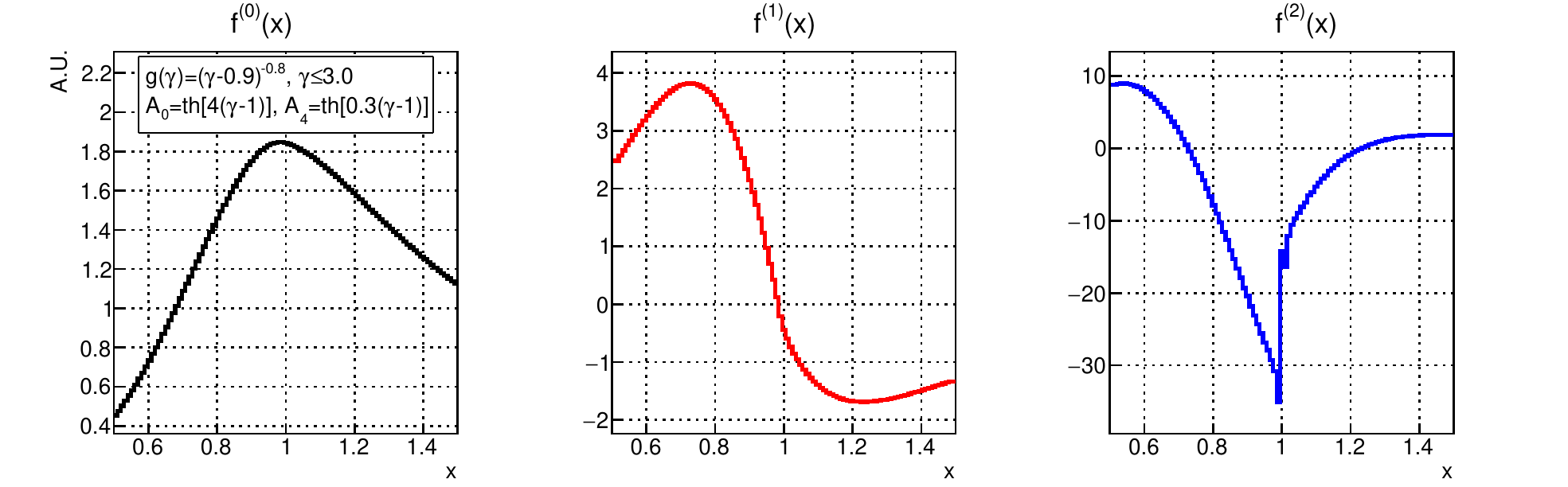}\\
   \caption{The lepton energy p.d.f $f\equiv f^{(0)}$ (left) of Eq.~\eqref{eq:diffxfull} with its first (center) and second (right) derivative for a boost p.d.f. of the form $g(\gamma)\propto(\gamma-0.9)^{-0.8}$ and the choice $(A_0,A_4)=$ $(0,0)$ (first row), $(2/3,0)$ (second row), $(0,1)$ (third row), and $(\mbox{tanh}\left[4(\gamma-1)\right], \mbox{tanh}\left[0.3(\gamma-1)\right])$ (fourth row). The decaying resonance is treated in the narrow-width approximation.}
   \label{fig:toy2}
 \end{center}
\end{figure}

\begin{figure}[hbtp]
 \begin{center}
   \includegraphics[width=1\textwidth]{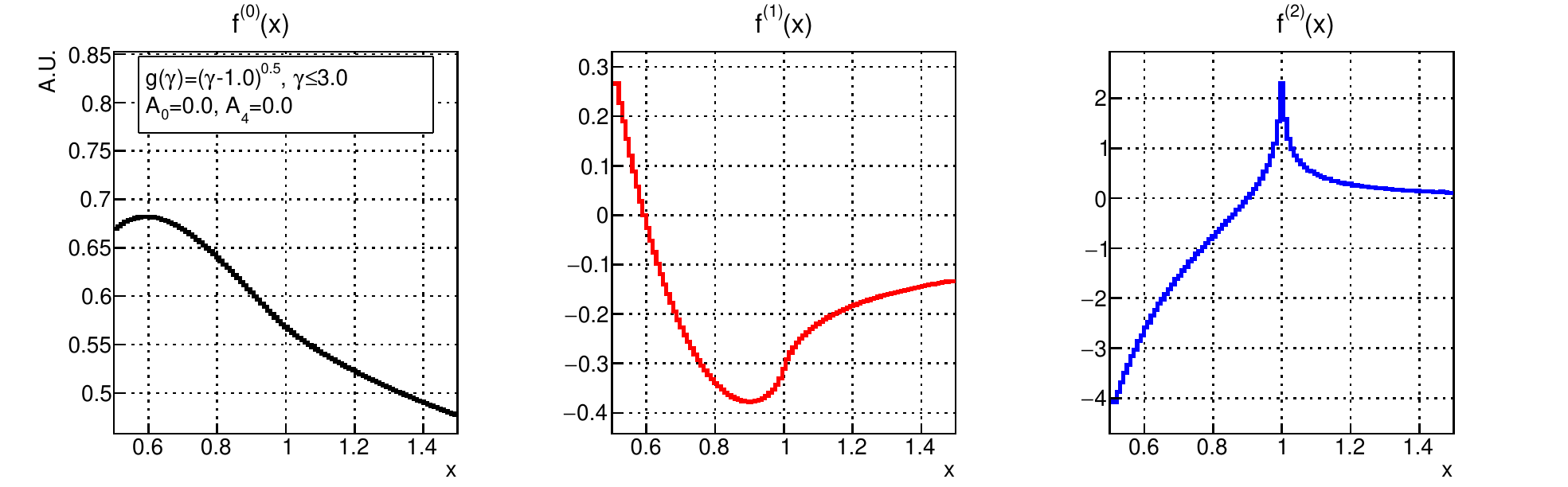}\\
   \includegraphics[width=1\textwidth]{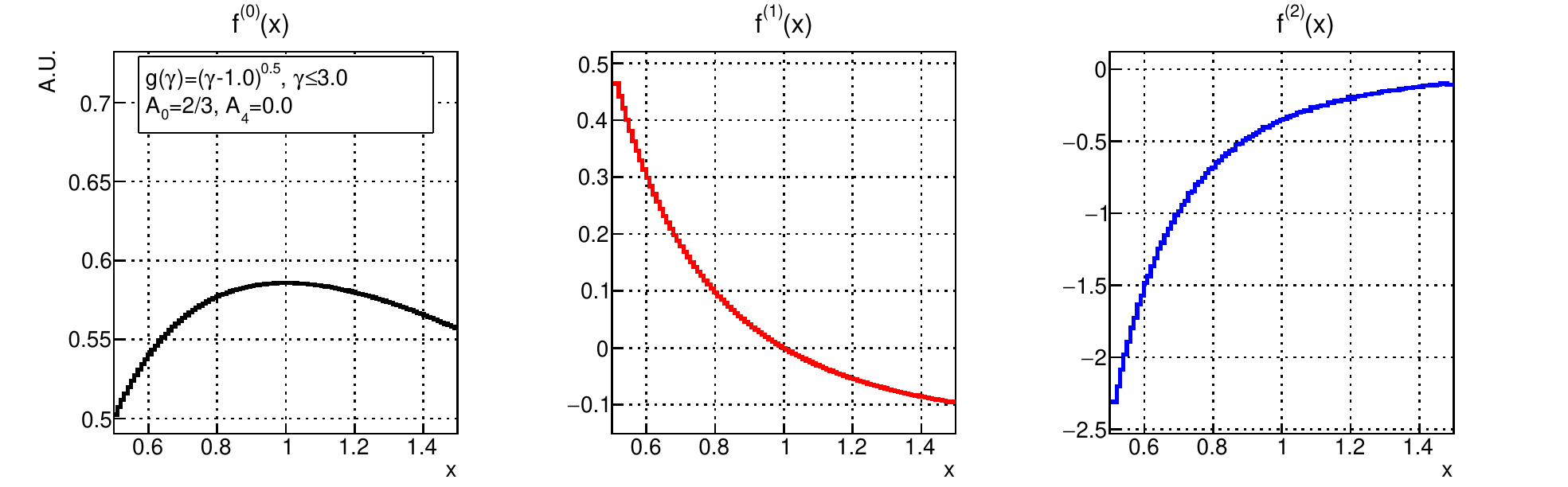}\\
   \includegraphics[width=1\textwidth]{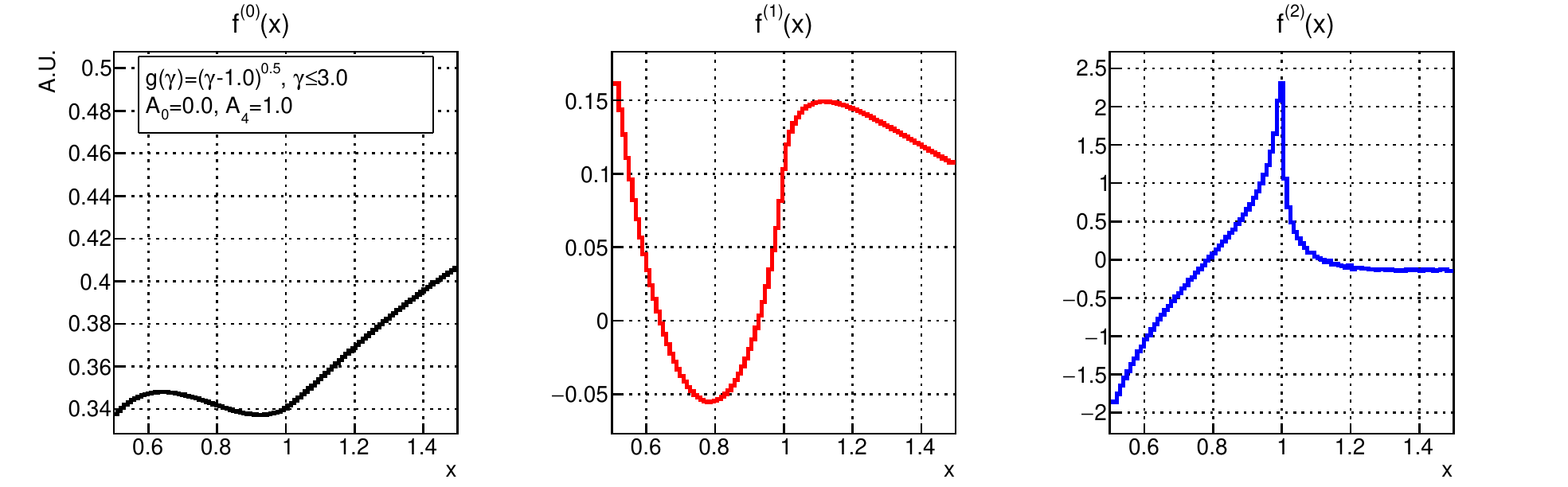}\\
   \includegraphics[width=1\textwidth]{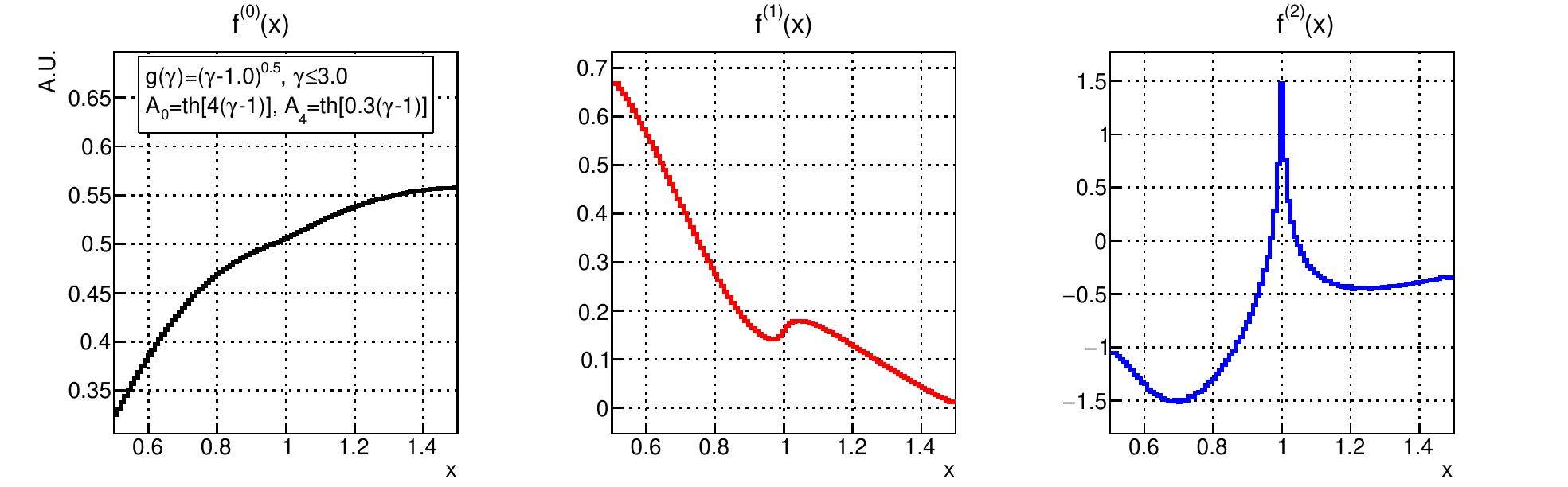}\\
   \caption{The lepton energy p.d.f $f \equiv f^{(0)}$ (left) of Eq.~\eqref{eq:diffxfull} with its first (center) and second (right) derivative for a boost p.d.f. of the form $g(\gamma)\propto(\gamma-1)^{\frac12}$ and the choice $(A_0,A_4)=$ $(0,0)$ (first row), $(2/3,0)$ (second row), $(0,1)$ (third row), and $(\mbox{tanh}\left[4(\gamma-1)\right], \mbox{tanh}\left[0.3(\gamma-1)\right])$ (fourth row). The decaying resonance is treated in the narrow-width approximation.}
   \label{fig:toy3}
 \end{center}
\end{figure}

The predictions of Eq.~\eqref{eq:diffxdelta2} have been verified numerically for selected choices of the functions $g$, $A_{0}$, and $A_{4}$. The three following functional forms for the boost factor p.d.f. have been studied:
\begin{itemize}
\item $g_{\rm exp}(\gamma) \propto (\gamma-1)\exp^{-(\gamma-1)}$. This function is analytic in $\gamma=1$, and is chosen as the prototype of a p.d.f. with $g^{(0)}=0$.
\item $g_{\rm pow}(\gamma)\propto(\gamma-0.9)^{-0.8}$. This function is analytic in $\gamma=1$, but this time $g^{(0)}>0$. The numerical values of the coefficients are somehow tuned on the empirical boost distribution for $W$ bosons production at the LHC when $\gamma-1$ is in excess of about $10^{-3}$, see Sec.~\ref{sec:WLHC}.
\item $g_{\rm sqrt}(\gamma)\propto(\gamma-1)^{\frac12}$. This function is chosen as the prototype of a p.d.f which is not analytic in $\gamma=1$. In particular, it is finite for $\gamma\to1^+$, but its first derivative is infinite.
\end{itemize}
For sake of numerical precision, the integration over the boost factors is restricted to the range $\gamma \in [1,3]$.
Both $g_{\rm exp}$ and $g_{\rm pow}$ are not integrable, so strictly speaking they cannot be interpreted as probability density functions. However, they can still provide a good approximation of physical densities for small values of $\gamma$.
For all three functions, the following values for $(A_0,A_4)$ have been studied:
\begin{enumerate}
\item  $(0,0)$, corresponding to an equal left and right polarization;
\item   $(\frac23,0)$, corresponding to an unpolarized resonance;
\item   $(0,1)$, corresponding to partial transverse polarization;
\item   $(\mbox{tanh}\left[4(\gamma-1)\right], \mbox{tanh}\left[0.3(\gamma-1)]\right)$, corresponding to a resonance which has equal left and right polarization at rest, and then it acquires both a longitudinal and a transverse polarization as the boost factor increases. The choice of numerical constants is somehow inspired by the case of $W$ production as discussed in Sec.~\ref{sec:WLHC}.
\end{enumerate}
The resulting probability density functions $f^{(0)}$ are shown in Fig.~\ref{fig:toy1}-\ref{fig:toy3} together with their first $f^{(1)}$ and second $f^{(2)}$ derivatives. The latter are estimated from finite differences of $f^{(0)}$ over an equally-spaced mesh of $x_i$ values:
\begin{subequations}\label{eq:derivatives}
\begin{align} 
\label{eq:derivatives:1}
f^{(1)}(x_i) & \approx \frac{f^{(0)}(x_i+d)-f^{(0)}(x_i-d)}{2d} \\
\label{eq:derivatives:2}
f^{(2)}(x_i) & \approx \frac{f^{(0)}(x_i+d)+f^{(0)}(x_i-d)-2f^{(0)}(x_i)}{d^2}
\end{align} 
\end{subequations}
where $d$ is the mesh size.

Figure~\ref{fig:toy1} shows the results for $g_{\rm exp}$ for each choice of the angular coefficients. Since $g^{(0)}=0$, we have $C=E=H=0$, as for Eq.~\eqref{eq:diffxdelta3}. Apart from case $2)$, where $x=1$ is also a local maximum of $f^{(0)}$, the first derivative does not vanish in general at $x=1$. However, the presence of a term like $|\epsilon^3|$ in Eq.~\eqref{eq:diffxdelta2} induces the presence of a cusp in the second derivative.

For $g_{\rm pow}$, which has $g^{(0)}>0$, additional sources of non-analyticity are present in Eq.~\eqref{eq:diffxdelta2}, clearly visible in Figure~\ref{fig:toy2}. In case $1)$, the Taylor expansion of $f^{(0)}$ contains a term of the form $E\epsilon|\epsilon|$ with $E\neq 0$, hence the second derivative receives a contribution from a step-function centered at $x=1$. In case $2)$, $C\neq0$ so that $x=1$ is a cusp: the first and second derivatives are thus locally proportional to a step-function and a delta function, respectively. In case $3)$, $H<0$, so that the first order diverges to $+\infty$ like $\ln|\epsilon|$ when $\epsilon\to 0$, whereas the second order derivative goes like $1/\epsilon$. Case $4)$ is qualitatively similar to the first.

Whenever $g$ or any of the two angular coefficients are not analytic at $\gamma=1$, like for $g_{\rm sqrt}$, Eq.~\eqref{eq:diffxdelta2} does not apply. The general appearance of step functions and poles in $x=1$ is however unchanged, as shown by Figure~\ref{fig:toy3}. In particular, a term of the form $\epsilon^2\ln|\epsilon|$ stems from the last but one line of Eq.~\eqref{eq:diffxdelta}.
The choice of $g_{\rm sqrt}$ is, however, special since, for an unpolarized resonance, it provides an analytic p.d.f.:
\begin{align} \label{eq:diffxdeltaNonAna}
f(1+\epsilon) \propto \int_{{\epsilon^2}/{2}}^{\delta}d\kappa \, & \kappa^{\frac12} \left[ 2^{-\frac12}  \kappa^{-\frac12} \right] + \int_{\delta}^{+\infty} d\kappa \, \left[ \hdots \right] = A + C\epsilon^2 + \hdots
\end{align}
In this last case, the mass estimator would be provided by $\mbox{argmax} [f]$.

\section{$W$ bosons at the LHC}\label{sec:W}

A special case of the problem studied in Sec.~\ref{sec:kin} is represented by $W$ bosons produced in hadron-hadron collisions and decaying into a lepton-neutrino pair. For the purpose of studying this particular process, a sample of proton-proton collision events at $\sqrt{s}=13$ TeV simulating the $pp\to W^\pm X$, $W^\pm \to \mu^\pm \nu_\mu$ reaction has been generated with NLO QCD accuracy using the \texttt{MG5\_aMC@NLO}~\cite{amc} event generator interfaced with {\tt Pythia8} for the parton shower~\cite{pythia}. The {\tt NNPDF3.0}~\cite{nnpdf} set is used to simulate the proton PDFs. A total of about $84$ millions of events are generated, with a fraction of negative weights such that the effective number of events is reduced by roughly a factor of two compared to the case of unweighted events. Given that the cross section for $W \to \mu \nu_\mu$ production is about $20.5$ nb at $\sqrt{s}=13$ TeV~\cite{FEWZ}, the simulated sample used for this study has the same statistical power of a sample of collision events corresponding to $1.9$ fb$^{-1}$ of integrated luminosity.

The natural width of the $W$ boson is $\Gamma_W\approx 2.08$ GeV~\cite{PDG}, corresponding to a value $\Delta\approx10^{-2}$ in Eq.~\eqref{eq:BW}. This is not negligible on the scale of a high-precision measurement of $M_W$, which targets a relative accuracy on the mass as low as $10^{-4}$~\cite{WMassCDF, WMassD0, WMassATLAS}. Hence, an ultimate calibration of the estimator is required to meet this level of accuracy.
In Sec.~\ref{sec:WLHC}, it was found that the boost factor p.d.f. for $W$ bosons produced in proton-proton collisions can be roughly approximated by a power law $g\sim (\gamma-1)^{\alpha}$: for small values of $\gamma-1$, i.e. $\lesssim 5\times 10^{-4}$, we have $\alpha\approx0.5$ and the $W$ boson is almost unpolarized in the helicity frame; for higher boost values, $\alpha\approx-0.8$, and a net polarization is built, ultimately dominated by a particular transverse mode. From the numerical simulations of Fig.~\ref{fig:toy2}-\ref{fig:toy3}, we could thus expect $x=1$ to be a local minimum of $f^{(2)}$. Indeed, the rising edge of $g$ populates only the region $|x-1|\lesssim 5\times 10^{-4}$, where it provides a smooth function $f^{(2)}$, similarly to the rightmost panel in the second row of Fig.~\ref{fig:toy3}. For larger boosts, $f^{(0)}$ should resemble more closely the plots in Fig.~\ref{fig:toy2}, featuring a deep minimum of $f^{(2)}$ at $x=1$. The whole picture is then smeared by the finite width of the $W$ boson.

\begin{figure}[hbtp]
 \begin{center}
   \includegraphics[width=1\textwidth]{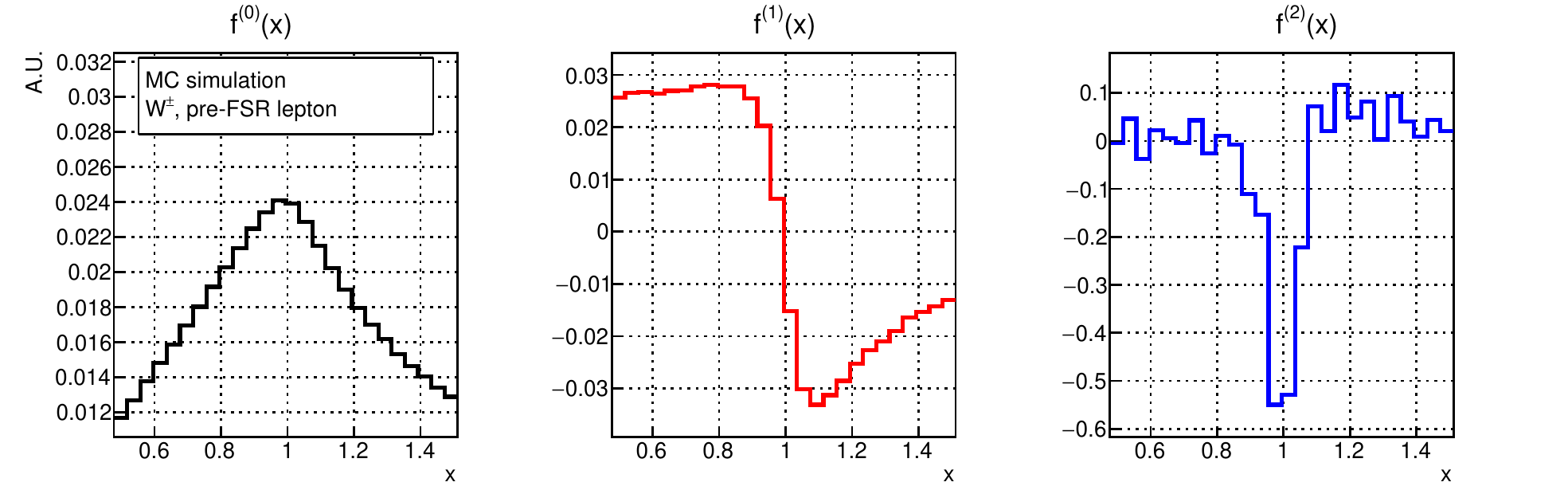}
   \caption{The p.d.f $f\equiv f^{(0)}$ from a simulation of $pp\to W^\pm X$, $W^\pm \to \mu^\pm \nu_\mu$ events (left) with its first (center) and second (right) derivative. In the right-hand panel, the visible bin-by-bin fluctuations are due to the limited size of the simulated sample.}
   \label{fig:toyMC}
 \end{center}
\end{figure}

Figure~\ref{fig:toyMC} shows the binned density $f^{(0)}$ obtained from the simulated sample of events. The first and second derivatives are estimated bin-wise in the same fashion as Eq.~\eqref{eq:derivatives}. A deep minimum in the histogram of $f^{(2)}$ at $x$ values close to unity is clearly visible. Interestingly, $x=1$ is also close to be a global maximum of $f^{(0)}$, a result that qualitatively recalls the last toy example in Fig.~\ref{fig:toy2}, where the boost p.d.f. and the angular coefficients were indeed tuned on the values extracted from the Monte Carlo simulation. 

\subsection{Detector acceptance and final-state radiation}\label{sec:detacc}

For the case of $W$ boson production and decay at the LHC, two effects break the mathematical hypotheses assumed to derive Eq.~\eqref{eq:diffx}: the presence of acceptance selection requirements, which are unavoidable in experiments at hadron colliders, and the emission of final-state photon radiation (FSR) from the charged lepton. Both affect the center-of-mass dynamics, albeit in different ways as discussed below.

When the detector coverage is incomplete, the harmonic polynomials that depend on $\phi^*$, which are themselves proportional to $\sin\theta^*$ and $\sin2\theta^*$~\cite{Mirkes}, don't average exactly to zero in some regions of the phase-space, thus adding spurious terms to the $c^*$ expansion of the decay angle distribution. Furthermore, the detector acceptance requirements, being based only on the kinematics of the visible decay products, affect the lepton reconstruction efficiency differently depending on the kinematics of the $W$ boson. The overall result is to modify the angular coefficients by boost-dependent efficiency factors $\rho_{0,4}(x \, | \, \gamma)$ in Eq.~\eqref{eq:diffxfull}, which are in general non-trivial functions of $x$. Here, we will study the effect of selection requirements realistic for general-purpose experiments like ATLAS~\cite{ATLAS} and CMS~\cite{CMS}, namely $|\eta|\leq2.5$ and $p_{\rm T}\geq25$ GeV, where $\eta$ and $p_{\rm T}$ are the muon pseudorapidity and transverse momentum, respectively. In the MC simulation we find these cuts to have an efficiency of about $77\%$ for $W^+$ and $84\%$ for $W^-$ for events with lepton energy $E\approx M_W/2$. 

The emission of FSR by the charged lepton perturbs the center-of-mass dynamics. The overall effect of such perturbation can be thought of as the convolution of the original harmonic polynomials with a smearing kernel, which introduces infinite harmonics in $c^*$. Furthermore, the emission of extra particles (photons and lepton-pairs) reduces the center-of-mass energy available for the muon and thus primarily affects the visible energy spectrum by an overall downward shift. This process is well-known~\cite{Calame} though, so that it could be in principle unfolded at the detector level to recover a pure QCD description of the final-state kinematics. For the purpose of studying this process, we will consider both an unrealistic scenario, where the charged leptons do not undergo photon radiation (pre-FSR leptons), and a realistic scenario where a QED-shower of the muons is simulated by the \texttt{Pythia8}  MC (bare leptons). 

\subsection{The search for a stationary point}

We now consider the problem of finding the stationary points of the higher-order derivatives of $f$. The rightmost histogram in Fig.~\ref{fig:toyMC} clearly shows that a local minimum of $f^{(2)}$ is present at $x\approx1$. The estimator of such point is, however, not uniquely defined. We won't address here the problem of finding the statistically optimal of such estimators. Instead, we decide to define the estimator implicitly as the root of a conveniently chosen function of the data. To this purpose, we first approximate the density $f$ with a polynomial function of degree $D$ centered at $x=1$:
\begin{align} \label{eq:polx}
f(x) \equiv f^{(0)}(x) \approx \sum_{n=0}^Dc_n(x-1)^n.
\end{align}
The coefficients $c_n$ in Eq.~\eqref{eq:polx} are determined from a fit to the simulated data by means of an analytic $\chi^2$ method. We then define the stationary points of the $i$-th order derivative implicitly as the roots of the $(i+1)$-th order derivative. The latter are determined numerically by using Halley's root-search method~\cite{Halley}, a variant of the classical Newton method. Statistical uncertainties on the coefficients of the polynomial fit are propagated to the roots $\hat{x}_i$ by means of pseudo-data, resulting in 68\% confidence level (CL) intervals.
This approach has a twofold advantage: it regularizes the statistical bin-by-bin fluctuations by the use of smooth functions and it allows for an analytic evaluation of the derivatives at any point $x$.

The energy spectrum is provided as a histogram with 100 MeV large bins. The central value of each bin is normalized to the constant $E_W=M_W/2$ to yield the dimensionless variable $x$. The fit is performed in the interval $E\in[36.2,44.3]$ GeV, corresponding to invariant masses of the $W$ bosons in a window of about $\pm4 \Gamma_W$ around $M_W$. Such range is large enough to provide acceptable fits with $D=4$, which is the minimum degree to define a unique root of the third derivative $\hat{x}_3$. 
We notice that this way of estimating the roots $\hat{x}_i$ is quite sensitive to border effects related to the choice of the fit range: since Eq.~\eqref{eq:polx} is only a local approximation of the density, discrepancies between the true spectrum and $f^{(0)}$ at the edges of the fit range tend to pull more strongly the coefficients associated with the large powers of $n$, which in turn affect more strongly the roots of the higher order derivatives. The bias associated with the choice of fit range will be eventually reabsorbed as part of the calibration procedure discussed in the next section. 

\begin{figure}[hbtp]
 \begin{center}
   \includegraphics[width=0.45\textwidth]{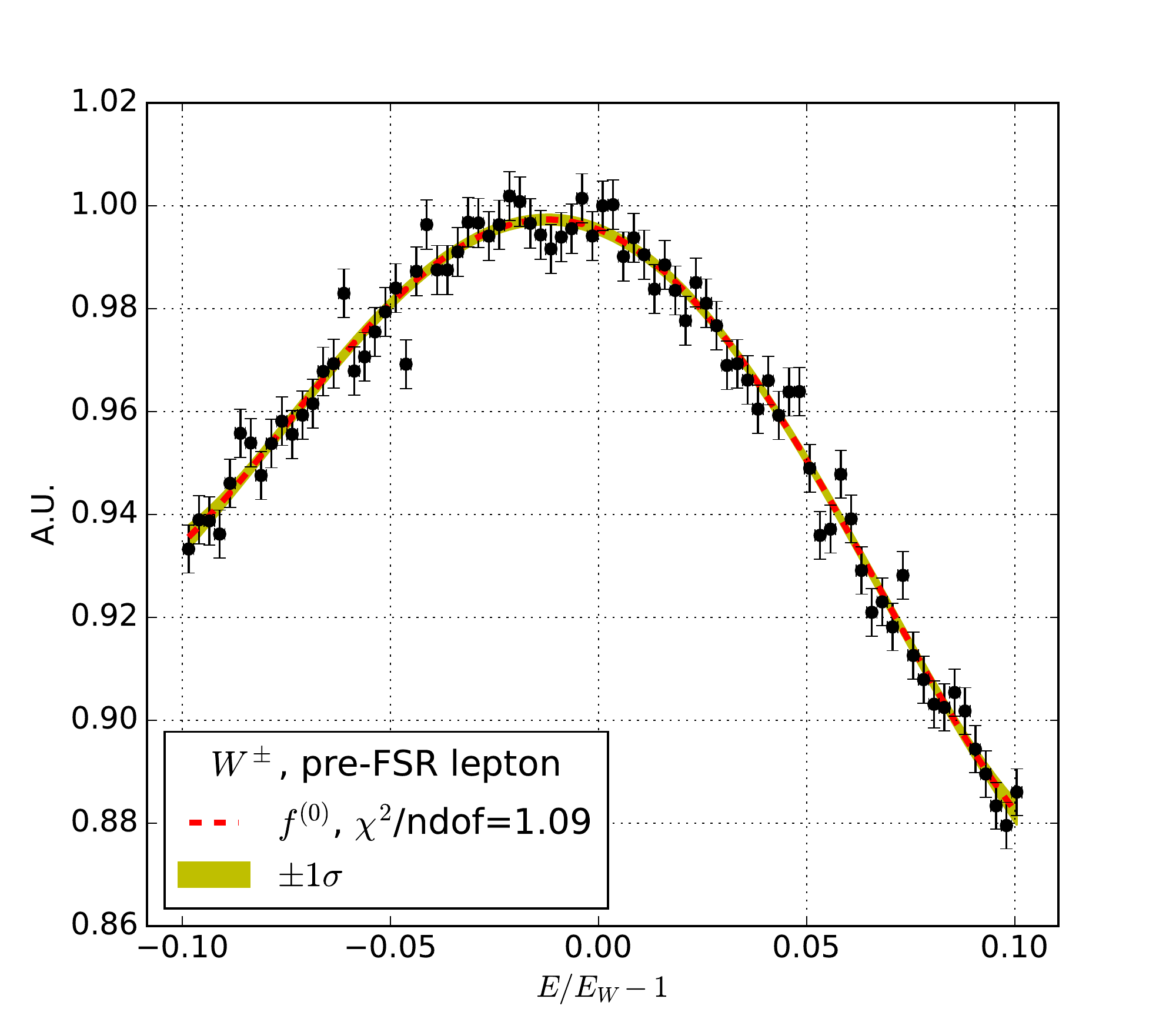}
   \includegraphics[width=0.45\textwidth]{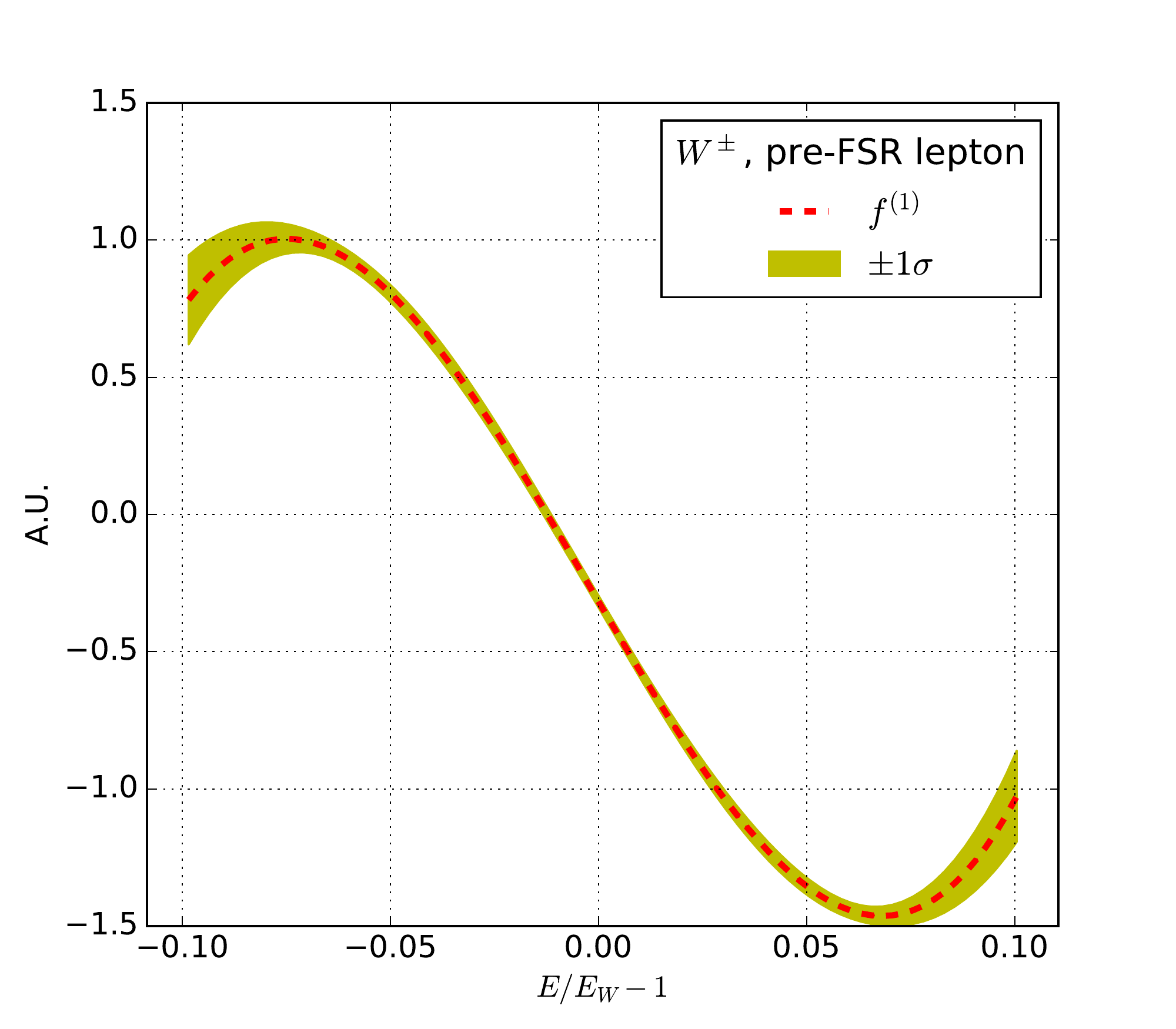}
   \includegraphics[width=0.45\textwidth]{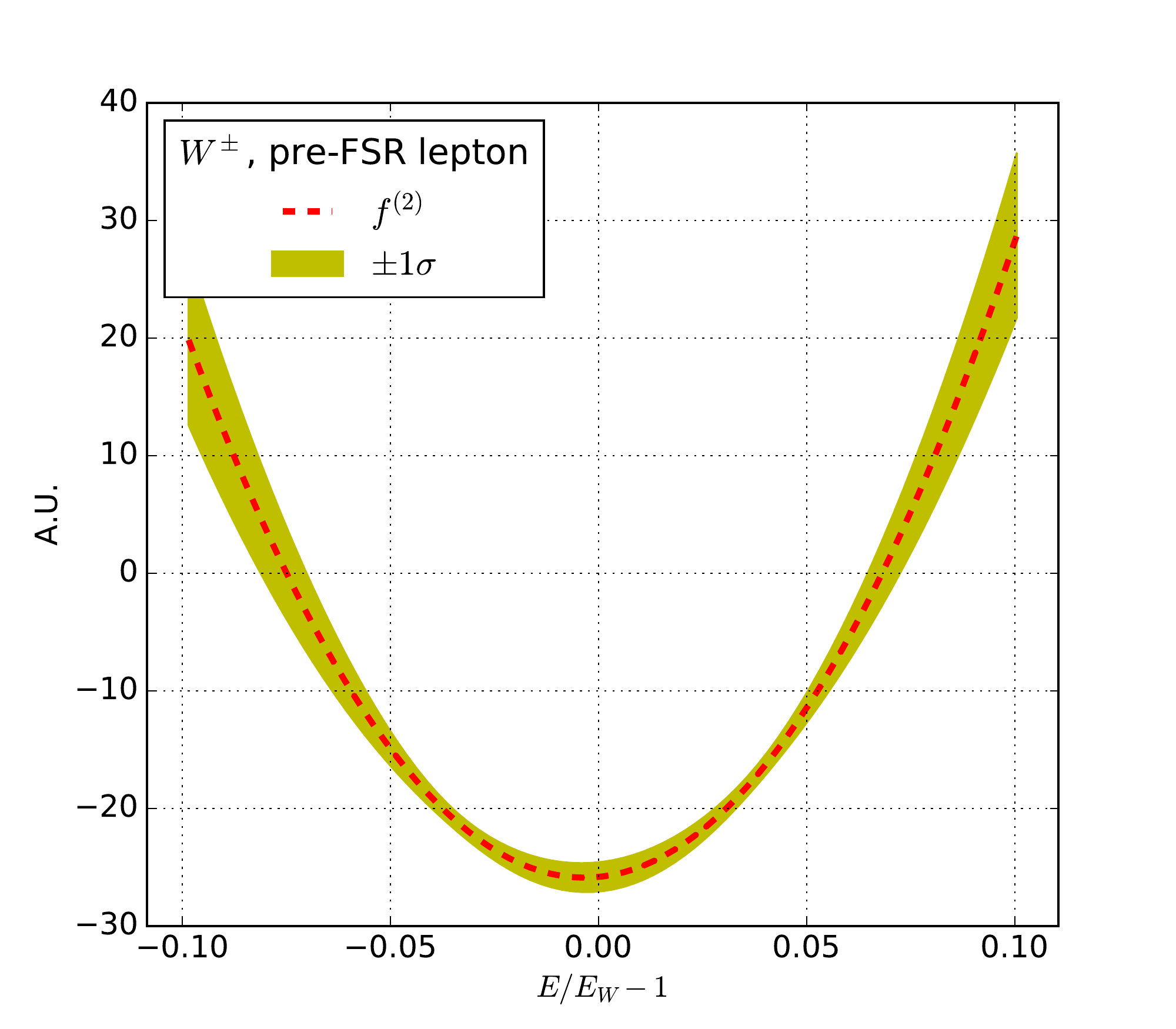}
   \includegraphics[width=0.45\textwidth]{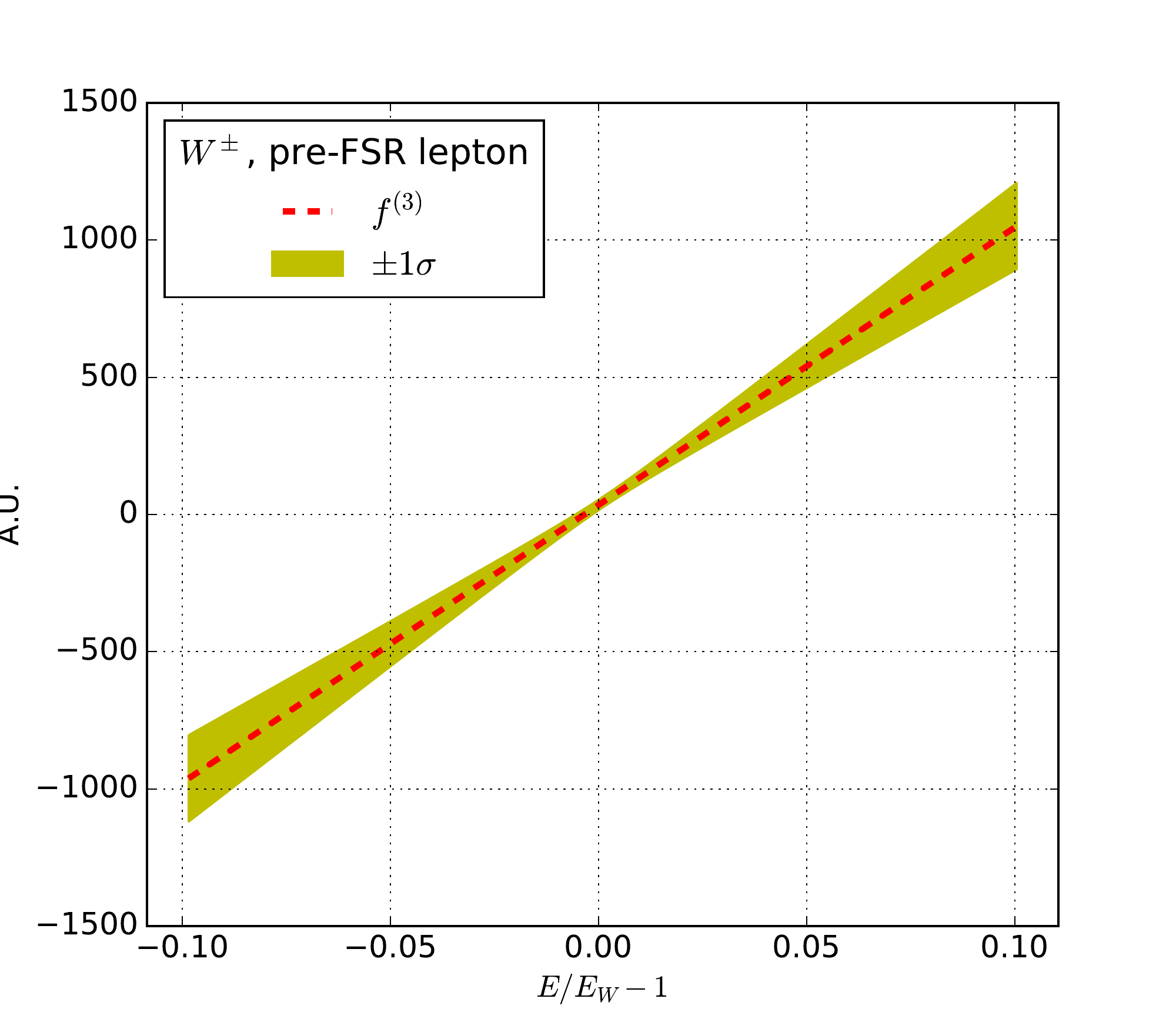}
   \caption{Polynomial fits to the distribution of the variable $x=E/E_W-1$, where $E$ is the lepton energy in the laboratory frame and $E_W=M_W/2$, obtained from a Monte Carlo simulation of the process $pp\to W^\pm X$, $W^\pm\to \mu^\pm\nu_\mu$. The distribution is fitted to a fourth-order polynomial (red dahsed-line). The solid area shows the 68\% CL interval as obtained from the covariance matrix of the fit. The first, second, and third derivatives of the fitted function are also shown with their uncertainty bands.}
   \label{fig:MCboth_full} 
 \end{center}
\end{figure}

For the sake of comparison, the root of the first derivative $\hat{x}_1$, which corresponds to a local maximum of $f^{(0)}$, is also studied. Positive and negative muon events are first considered separately. Since the two samples of events provide consistent results, they are ultimately combined to maximize the statistical accuracy of the analysis. 
The result of the fit to the simulated data is shown in Fig.~\ref{fig:MCboth_full}, together with the first, second, and third derivative of the fitted polynomial function.
As expected, the second derivative features a local minimum around $x=1$ identified by the root $\hat{x}_3$ of the third derivative. 

\subsection{Calibration curve}

\begin{figure}[hbtp]
 \begin{center}
   \includegraphics[width=0.45\textwidth]{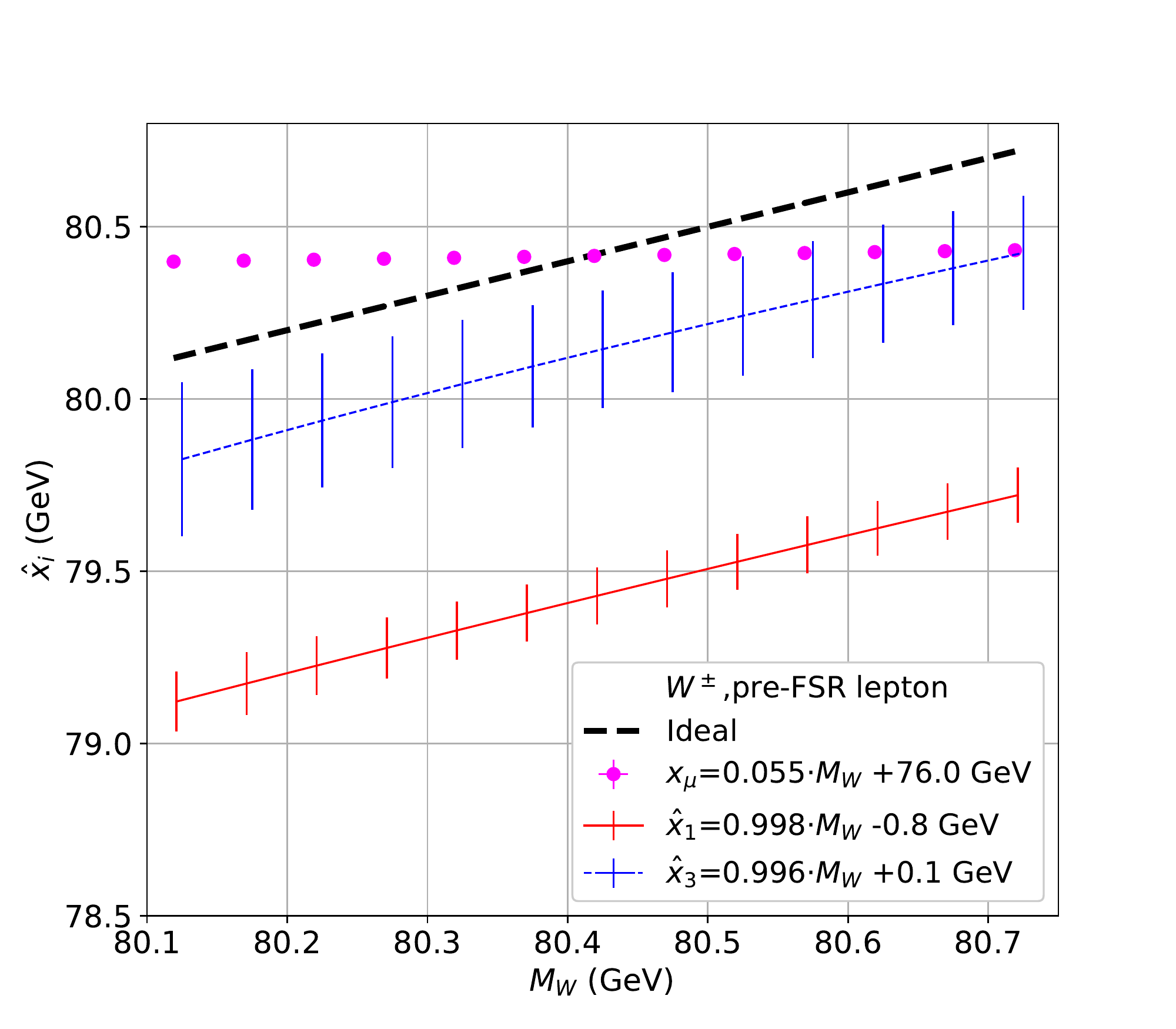}
   \includegraphics[width=0.45\textwidth]{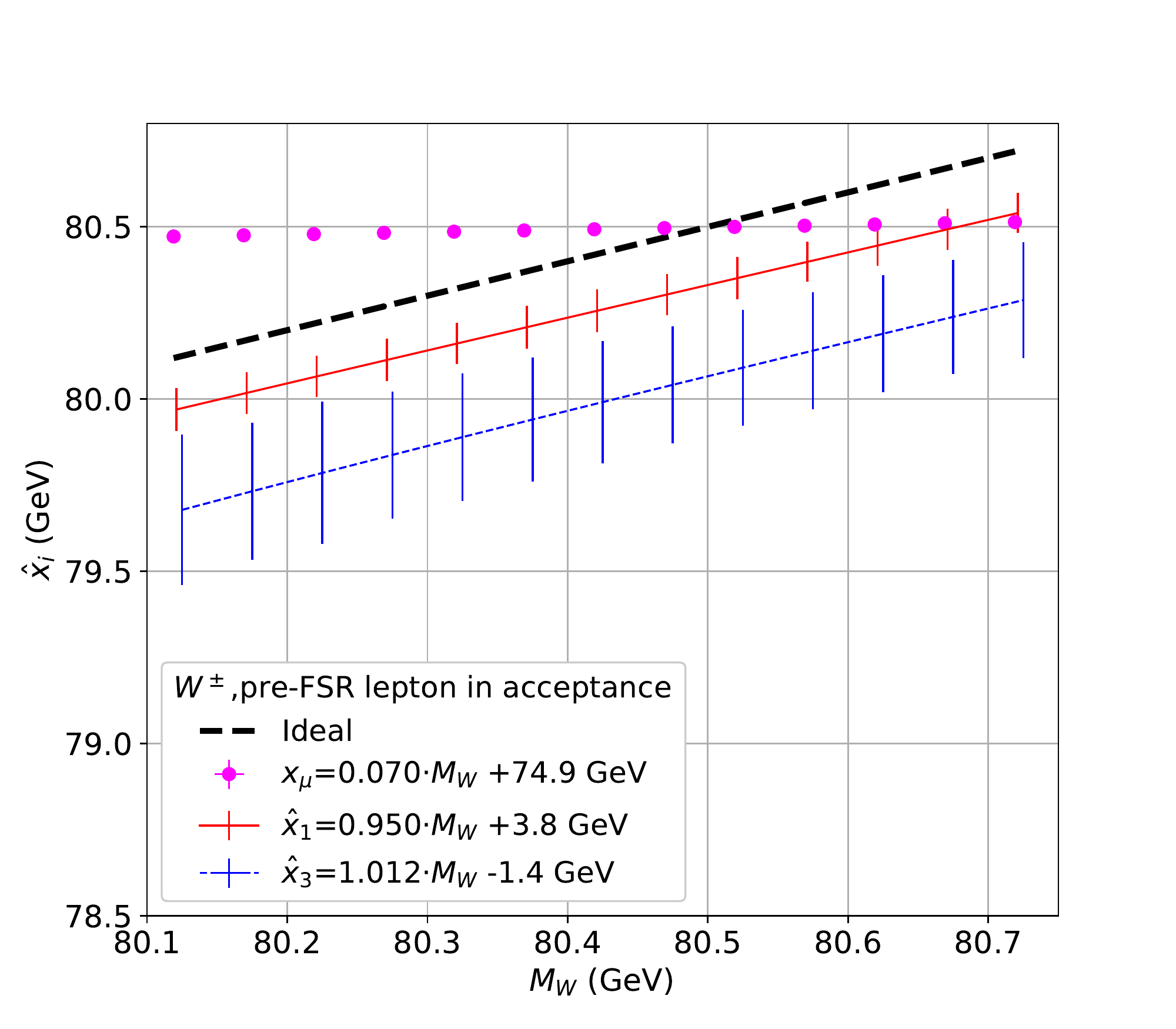}\\
   \caption{The roots $\hat{x}_1$ and $\hat{x}_3$ as a function of $M_W$ obtained from a fit to a MC simulated sample of $pp\to W^\pm X$, $W^\pm \to \mu^\pm \nu_\mu$ events, where pre-FSR muons are considered in the full phase-space (left) or within the detector acceptance (right). For comparison, the mean value $x_\mu$ of the distribution in the same range of the fit is also shown.}
   \label{fig:calibration_full_preFSR}
 \end{center}
\end{figure}

The calibration of the $\hat{x}_i$ estimator is determined by reweighting the same MC sample to different values of $M_W$. The fit is then repeated for each mass-reweighted sample and new roots $\hat{x}_i$ are computed, resulting in a calibration curve $\hat{M}_W=\hat{M}_W(\hat{x}_i)$. Figure~\ref{fig:calibration_full_preFSR} shows such curves separately for muons in the full phase-space and for muons within the detector acceptance as defined in Sec.~\ref{sec:detacc}. The points $\hat{x}_i$ are then interpolated through a linear regression.

The response of $\hat{x}_3$ to a change of $M_W$ is found to be linear to better than 1\%. This fact is reassuring and confirms that $\hat{x}_3$ is indeed a good estimator of $M_W$. For comparison, the root of the first derivative $\hat{x}_1$, and the mean value $x_\mu$ in the same range of $x$ values considered in the fit, are also reported as a function of $M_W$. The former is found to have a good linear response but a larger offset compared to $\hat{x}_3$ (800 MeV against 100 MeV). Instead, the mean value $x_\mu$ is very mildly related to $M_W$, which makes it a rather poor estimator of the mass. This is however mainly an artifact of considering a narrow range of $x$ values: as illustrated by the first panel of Fig.~\ref{fig:MCboth_full}, the function $f^{(0)}$ in the neighborhood of $x=1$ is a concave function: a tiny shift $\delta x$ of the peak position does not change the mean of the distribution to first-order in $\delta x$. It is also interesting to study the response of the three estimators to a restriction of the lepton phase-space. This is shown in the right panel of Fig.~\ref{fig:calibration_full_preFSR}. Both ${x}_\mu$ and $\hat{x}_1$ are found to be significantly affected by acceptance selection requirements, i.e. their values change compared to the full-acceptance case by more than their statistical uncertainty. On the contrary, $\hat{x}_3$ is more stable, changing by less than one standard deviation.

\begin{figure}[hbtp]
 \begin{center}
   \includegraphics[width=0.45\textwidth]{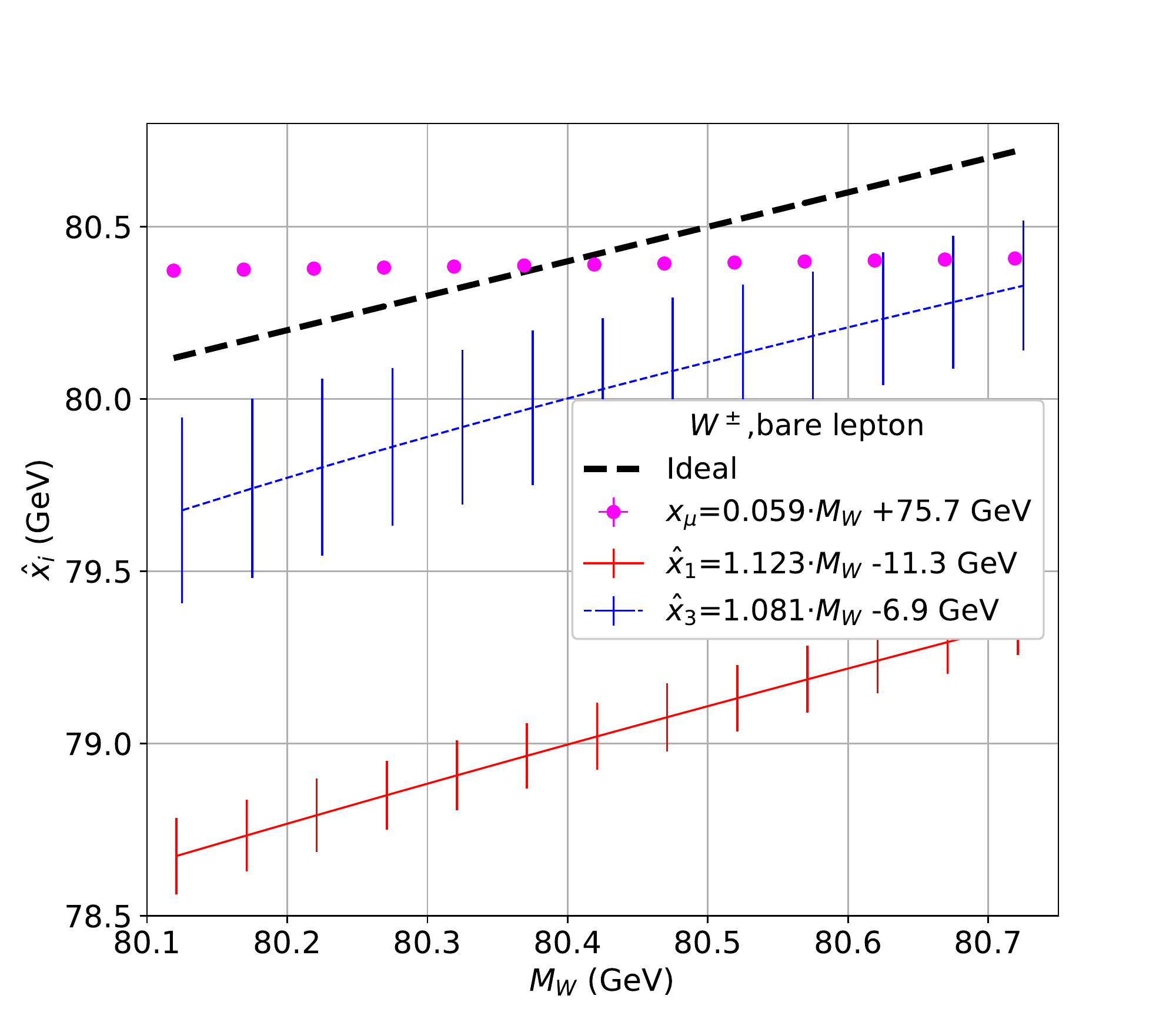}
   \includegraphics[width=0.45\textwidth]{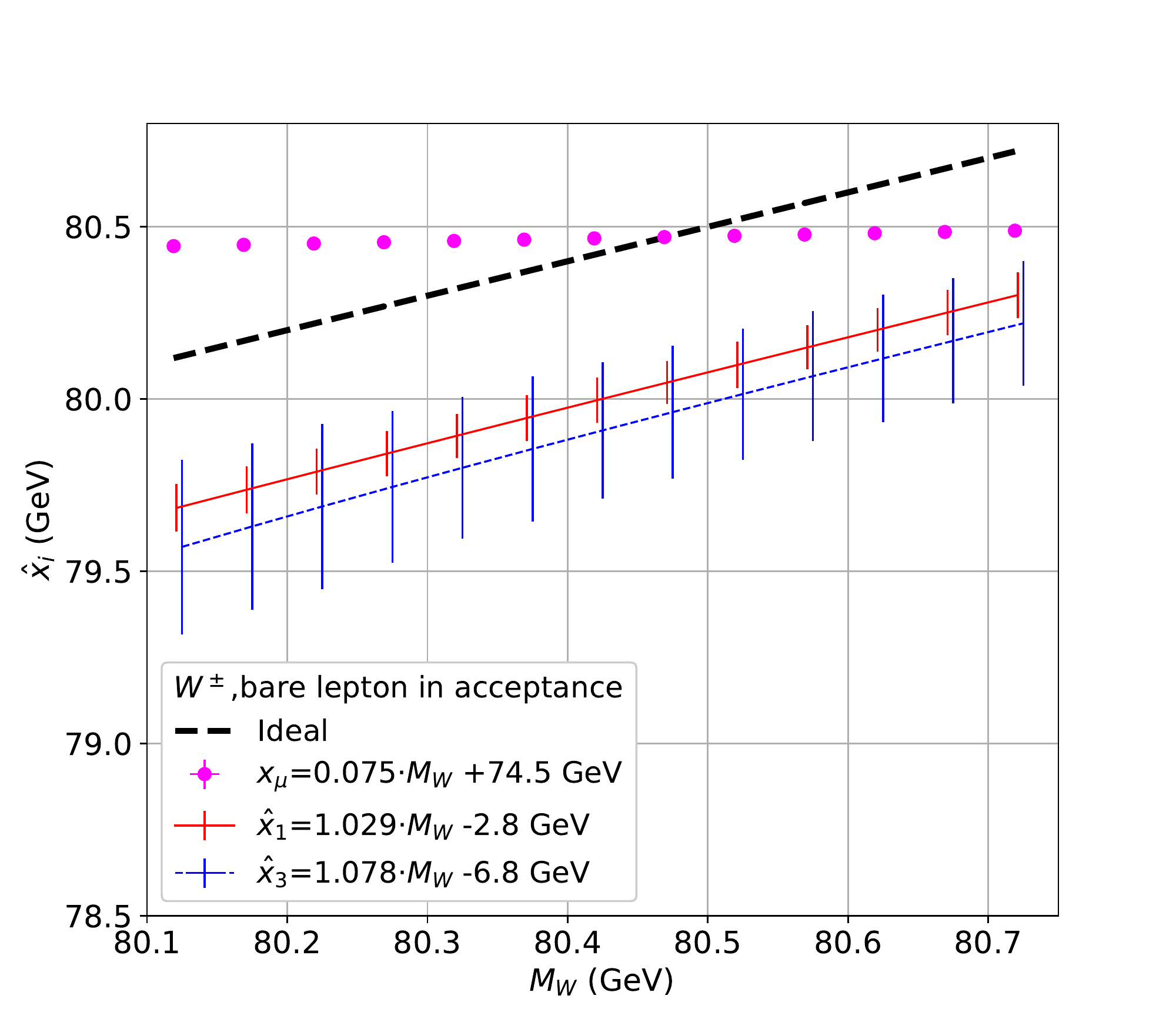}\\
   \caption{The roots $\hat{x}_1$ and $\hat{x}_3$ as a function of $M_W$ obtained from a fit to a MC simulated sample of $pp\to W^\pm X$, $W^\pm\to \mu^\pm\nu_\mu$ events, where bare muons are considered in the full phase-space (left) or within the detector acceptance (right). For comparison, the mean value $x_\mu$ of the distribution in the same range of the fit is also shown.}
   \label{fig:calibration_full_bare}
 \end{center}
\end{figure}

Finally, an identical analysis is repeated considering bare leptons instead of pre-FSR leptons. The results are shown in Fig.~\ref{fig:calibration_full_bare}. Besides an overall shift of about $200$ MeV, which can be ascribed to the loss of energy drained away by FSR~\cite{Calame}, the linear response of $\hat{x}_{1}$ and $\hat{x}_{3}$ to $M_W$ is found to be preserved.

\subsection{Residual model dependence}

The residual model-dependence of $\hat{x}_i$ on the production and decay dynamics will be eventually incorporated as a systematic uncertainty on the calibration curve. For the purpose of assessing the level of such relic model-dependence, the same MC sample has been reweighted to nine different sets of values for the renormalization and factorization scales ($\mu_R$, $\mu_F$)~\cite{amc} and to one-hundred MC replicas of the same PDF fit~\cite{nnpdf}. In the first case, the maximal deviation of the roots compared to the value obtained for the central choice of scales is symmetrized and used as a proxy of the perturbative scale uncertainty on $\hat{x}_i$. In the second case, the RMS of the distribution of roots is considered as systematic uncertainty related to the imperfect knowledge of the PDFs.

The results are shown in Table~\ref{tab:shift}. The variance of the weights used to modify the simulated sample contributes to these uncertainties. The latter is assessed by means of pseudo-experiments where the data in the nominal sample are randomized to account for the extra uncertainty added by the reweighting. We find the additional statistical fluctuation introduced by the reweighting to contribute to $\hat{x}_{i}$ by an amount corresponding to about 30\% (10\%) of the Poisson uncertainty on the fitted value for $i=1$ ($3$). The reduction in statistical uncertainty after applying the acceptance selection requirements, most noticeable for $\hat{x}_1$, is due to the change of the functional form of the energy spectrum. Indeed, if the selection efficiency were independent of $E$, the effect of the acceptance requirements would be to enlarge $\sigma_{\rm stat}$ by some 10\% due to the reduced event yield. Instead, the acceptance selection efficiency increases with $E$: the net effect is to sculpt the energy spectrum in such a way that the curvature of $f^{(0)}$ around $x=1$ increases, thus reducing the statistical uncertainty on the position of the maximum.

We can now summarize the results of this study as follows:
\begin{enumerate}
\item The root $\hat{x}_3$ is less affected than $\hat{x}_1$ by changes in the modeling of $W$ production and decay dynamics induced by different choices of perturbative scales and PDFs.
\item The root $\hat{x}_1$ shows a significant systematic uncertainty, i.e. larger than the Poisson fluctuation introduced by the reweighting. This corroborates the observation that $\hat{x}_1$ depends more than $\hat{x}_3$ on the underlying dynamics, as also deduced by its larger sensitivity to the acceptance requirements.
\item The residual scale and PDF uncertainty on $\hat{x}_3$ is consistent with the Poisson fluctuation introduced by the reweighting; within the statistical accuracy of this study, there is no indications of residual systematic bias.
\item A statistical-only uncertainty on $\hat{x}_3$ corresponding to $15$ MeV uncertainty on $M_W$ could be reached with about 300 fb$^{-1}$ of LHC data, which is within the reach of the Run 3 of the LHC.
\end{enumerate}

\begin{table}[!htb]
\centering
\begin{tabular}{c  | c c c | c c c | c c c}
\hline
                     & \multicolumn{3}{c}{$W^{+}$} & \multicolumn{3}{c}{$W^{-}$}  & \multicolumn{3}{c}{$W^{\pm}$} \\
\hline
      & $\frac{\sigma_{\rm stat}}{\rm MeV}$  & $\frac{\sigma_{\rm scale}}{\sigma_{\rm stat}}$   & $\frac{\sigma_{\rm PDF}}{\sigma_{\rm stat}}$    & $\frac{\sigma_{\rm stat}}{\rm MeV}$  & $\frac{\sigma_{\rm scale}}{\sigma_{\rm stat}}$   & $\frac{\sigma_{\rm PDF}}{\sigma_{\rm stat}}$ & $\frac{\sigma_{\rm stat}}{\rm MeV}$  & $\frac{\sigma_{\rm scale}}{\sigma_{\rm stat}}$   & $\frac{\sigma_{\rm PDF}}{\sigma_{\rm stat}}$\\
\hline
$\hat{x}_1$         & 120            &  22\%       & 39\%     & 110             & 38\%        & 40\%     &  80          & 10\%        & 49\%\\
$\hat{x}_3$         & 230            &  5\%         & 6\%       & 370             & 18\%        &   4\%     & 180            & 17\%        & 6\%\\ 
\hline\hline
$\hat{x}_1$         & 80              &  14\%       & 44\%     & 90               & 41\%        & 39\%     &   60            & 31\%        & 52\%\\ 
$\hat{x}_3$         & 210            &  6\%       & 10\%      & 350             & 11\%        & 5\%     & 180            & 6\%        & 10\%\\ 
\hline
\end{tabular}
\caption{\label{tab:shift} The statistical uncertainty (in MeV) and the scale and PDF uncertainty on $\hat{x}_1$ and $\hat{x}_3$, separately for $W^+$, $W^-$, and their combination, without (top rows) or with (bottom rows) acceptance requirements, obtained from a Monte Carlo simulated sample of $pp\to W^\pm X$, $W^\pm \to \mu^\pm\nu_\mu$ events, corresponding to an integrated luminosity of $1.9$ fb$^{-1}$. The relative uncertainty on $\sigma_{\rm scale}$ and $\sigma_{\rm PDF}$ is about 30\% and 10\%, respectively. The relative statistical uncertainty on $\sigma_{\rm stat}$ is estimated using psuedo-experiments to be about 5\% and 25\% for $\hat{x}_1$ and $\hat{x}_2$, respectively.}
\end{table}

\subsection{Background}

The analysis discussed in the previous paragraphs neglects the presence of background sources. Since the measurement of a stationary point $\hat{x}_i$ relies on a local description of the energy p.d.f. in the neighborhood of $x=1$, any background with a non-flat p.d.f. has the potential to bias the mass estimator. For the case of $W$ boson production at hadron colliders~\cite{WMassCDF,WMassD0,WMassATLAS}, three major background processes should be considered: $i)$ multi-jet production, where the muon comes from hadron decays within a jet, $ii)$ Drell-Yan production of a muon pair, where one of the muons escapes detection, and $iii)$ top quark ($t$) production. The latter two cases are the least harmful. Indeed, neutral Drell-Yan events pass the event selection criteria to the extent that one of the muon is emitted with either soft $p_{\rm T}$ or large values of $|\eta|$. In turn, this condition preferentially selects events where the intermediate $Z/\gamma^*$ boson is produced with a finite boost. As for Eq.~\eqref{eq:diffxdelta2} with $g^{(k)}=0$, this implies that the $E$ spectrum of the selected muons in the neighborhood of $M_Z/2$ must have a flat second-order derivative. Similarly, muons in top quark events come from the decay of boosted $W$ bosons, since, in the rest frame the decaying $t$ quark, the $W$ boson recoils againts a $b$ quark with $|\bm p_W^*|\approx 0.4 \, m_t$.
For multi-jet production, these arguments do not hold and a detailed data-driven estimation of the functional form should be performed. However, we remark that the analysis discussed here is robust against changes in muon acceptance, as observed in Sec.~\ref{sec:detacc}. Since the multi-jet background is reducible by either tighter identification criteria on the muon or by stricter requirements on the missing energy or transverse mass in the event, we do expect room for optimization in case the functional form of this background were found to be measurable with only limited accuracy.

\subsection{Outlook}

A more refined analysis of the residual theoretical uncertainties would require the simulation of a much larger data sample and a careful treatment of other model effects, like non-perturbative physics, mixed QCD-QED corrections, etc. (see e.g. Ref.~\cite{Alioli} for a recent review). Similarly, experimental uncertainties from the backgrounds, the bias due to the choice of a fixed-order polynomial to fit the data, the impact of the lepton energy scale uncertainty, etc., should be thoroughly assessed. This is beyond the scope of this work. The study presented here confirms that a stationary point in the second derivative of the lepton energy density is a good estimator of $M_W$ and that it is robust against changes of the underlying $W$ boson production and decay dynamics, detector acceptance requirements, and the emission of photon radiation.

\section{Conclusions}

We have considered the two-fermion decay of a spin-1 resonance of mass $M$, and analyzed the lepton energy distribution in the laboratory frame in full generality, i.e. regardless of the underlying production and decay dynamics of the resonance. In particular, we have studied the analyticity of the probability density in the neighborhood of $M/2$. We find that the density at this point is not analytic for a narrow-width resonance. In particular, we have studied the conditions for which a singular point appears in the higher-order derivatives of the density, and found that the second derivative is likely to develop a cusp or a pole. Exact formulas have been derived under the assumption that the distribution of boost factors $\gamma$ and the polarization of the resonance are described by regular functions of $\gamma$. The formulas have been qualitatively validated with toy examples of production and decay of a narrow-width resonance. When a finite width of the resonance is accounted for, the regularity is restored such that cusps or poles are smoothed into local stationary points potentially displaced from $M/2$. The size of such displacement depends on the width of the resonance, but partially also on the production dynamics, thus requiring an ultimate calibration. The quest for stationary points in the higher-order derivatives of the energy density function is thus advocated as a way to estimate $M$ with possibly limited knowledge of the underlying production and decay dynamics of the resonance. A special case is represented by the production of $W$ bosons in proton-proton collisions, which has been studied on a Monte Carlo simulation of this process, assuming LHC-like conditions on the proton beams. As expected, a stationary point in the second derivative is found close to $M_W/2$. The robustness of this point as an estimator of $M_W$ has been studied by considering the effect of detector acceptance requirements, the emission of final-state radiation, changes of the perturbative calculation of scattering amplitudes for $W$ production, the proton PDFs, and the input $W$ boson mass. Interestingly, such a mass estimator features a good linearity, a small bias, and is rather resilient to changes in the lepton acceptance and in the modeling of the $W$ boson production dynamics. An ultimate assessment of the residual model-dependence is left for future work.

\acknowledgments
We would like to thank Roberto Franceschini for a fruitful discussion during the CMS Italian National Meeting held in Piacenza in 2017, which triggered this work. We are also indebted to Nico Kleijne for reading the manuscript and double-checking the formulas. 





\begin{thebibliography}{99}


\bibitem{James} F.~James, \textit{Statistical methods in experimental physics}, World Scientific, Hackensack USA (2006) p. 345 [\href{http://inspirehep.net/record/739783?ln=en}{{\tiny IN}SPIRE}].

\bibitem{WMassCDF} CDF Collaboration, \textit{Precise measurement of the W-boson mass with the CDF II detector}, \href{http://doi.org/10.1103/PhysRevLett.108.151803}{\textit{Phys. Rev. Lett.} \textbf{108} (2012) 151803} [\href{https://arxiv.org/abs/1203.0275}{\texttt{arXiv:1203.0275}}] [\href{https://inspirehep.net/search?p=find+eprint+1203.0275}{{\tiny IN}SPIRE}].

\bibitem{WMassD0}  D0 Collaboration, \textit{Measurement of the W boson mass with the D0 detector}, \href{http://doi.org/10.1103/PhysRevD.89.012005}{\textit{Phys. Rev. D} \textbf{89} (2014) 012005} [\href{https://arxiv.org/abs/1310.8628}{\texttt{arXiv:1310.8628}}] [\href{https://inspirehep.net/search?p=find+eprint+1310.8628}{{\tiny IN}SPIRE}].

\bibitem{WMassATLAS} ATLAS Collaboration, \textit{Measurement of the W-boson mass in pp collisions at $\sqrt{s}=7$ TeV with the ATLAS detector}, \href{http://doi.org/10.1140/epjc/s10052-017-5475-4}{\textit{Eur Phys. J. C} \textbf{77} (2018) 110} [\href{https://arxiv.org/abs/1701.07240}{\texttt{arXiv:1701.07240}}] [\href{https://inspirehep.net/search?p=find+eprint+1701.07240}{{\tiny IN}SPIRE}].

\bibitem{Kim} I.W.~Kim, \textit{Algebraic singularity method for mass measurements with missing energy}, \href{http://doi.org/10.1103/PhysRevLett.104.081601}{\textit{Phys. Rev. Lett.} \textbf{104} (2010) 081601} [\href{https://arxiv.org/abs/0910.1149}{\texttt{arXiv:0910.1149}}] [\href{https://inspirehep.net/search?p=find+eprint+0910.1149}{{\tiny IN}SPIRE}].

\bibitem{Han} T.~Han, I.W.~Kim and J.~Song, \textit{Kinematic cusps: determining the missing particle mass at colliders}, \href{https://doi.org/10.1016/j.physletb.2010.09.010}{\textit{Phys. Lett. B} \textbf{693} (2010) 575} [\href{https://arxiv.org/abs/0906.5009}{\texttt{arXiv:0906.5009}}] [\href{https://inspirehep.net/search?p=find+eprint+0906.5009}{{\tiny IN}SPIRE}].

\bibitem{Rujula} A.~De R\'ujula and A.~Galindo, \textit{Measuring the W-boson mass at a hadron collider: a study of phase-space singularity methods}, \href{http://doi.org/10.1007/JHEP08(2011)023}{\textit{JHEP} \textbf{08} (2011) 23} [\href{https://arxiv.org/abs/1106.0396}{\texttt{arXiv:1106.0396}}] [\href{https://inspirehep.net/search?p=find+eprint+1106.0396}{{\tiny IN}SPIRE}].

\bibitem{Franceschini} K.~Agashe, R.~Franceschini and D.~Kim, \textit{A simple, yet subtle invariance of two-body decay kinematics}, \href{http://doi.org/10.1103/PhysRevD.88.057701}{\textit{Phys. Rev.} D \textbf{88} (2013) 057701} [\href{https://arxiv.org/abs/1209.0772}{\texttt{arXiv:1209.0772}}] [\href{https://inspirehep.net/search?p=find+eprint+1209.0772}{{\tiny IN}SPIRE}].

\bibitem{CMS-top} CMS Collaboration, \textit{Measurement of the top-quark mass from the b jet energy spectrum}, \href{https://cds.cern.ch/record/2053086}{CMS-PAS-TOP-15-002} [\href{http://inspirehep.net/record/1393817?ln=en}{{\tiny IN}SPIRE}].

\bibitem{PDG} Particle Data Group, \textit{Review of Particle Physics}, \href{http://doi.org/10.1103/PhysRevD.98.030001}{\textit{Phys. Rev. D} \textbf{98} (2018) 030001} [\href{http://inspirehep.net/record/1688995}{{\tiny IN}SPIRE}]. 

\bibitem{Mirkes} E.~Mirkes, \textit{Angular decay distribution of leptons from W-bosons at NLO in hadronic collisions}, \href{https://doi.org/10.1016/0550-3213(92)90046-E}{\textit{Nucl. Phys. B} \textbf{387} (1992) 3} [\href{http://inspirehep.net/record/335604?ln=en}{{\tiny IN}SPIRE}].

\bibitem{Petronzio} G.~Parisi and R.~Petronzio, \textit{Small transverse momentum distributions in hard processes}, \href{https://doi.org/10.1016/0550-3213(79)90040-3}{\textit{Nucl. Phys. B} \textbf{154} (1979) 427} [\href{http://inspirehep.net/record/140188}{{\tiny IN}SPIRE}].

\bibitem{CS} J.C.~Collins and D.E.~Soper, \textit{Angular distribution of dileptons in high-energy hadron collisions}, \href{http://doi.org/10.1103/PhysRevD.16.2219}{\textit{Phys. Rev. D} \textbf{16} (1977) 2219} [\href{http://inspirehep.net/record/5163}{{\tiny IN}SPIRE}].

\bibitem{amc} J.~Alwall et al., \textit{The automated computation of tree-level and next-to-leading order differential cross sections, and their matching to parton shower simulations}, \href{http://doi.org/10.1007/JHEP07(2014)079}{\textit{JHEP} \textbf{07} (2014) 079} [\href{https://arxiv.org/abs/1405.0301}{\texttt{arXiv:1405.0301}}] [\href{https://inspirehep.net/search?p=find+eprint+1405.0301}{{\tiny IN}SPIRE}].

\bibitem{pythia} T.~Sjostrand, S.~Mrenna and P.~Skands, \textit{PYTHIA6.4 physics and manual}, \href{http://doi.org/10.1088/1126-6708/2006/05/026}{\textit{JHEP} \textbf{05} (2006) 026} [\href{https://arxiv.org/abs/hep-ph/0603175}{\texttt{hep-ph/0603175}}] [\href{https://inspirehep.net/search?p=find+eprint+hep-ph/0603175}{{\tiny IN}SPIRE}].

\bibitem{nnpdf} NNPDF Collaboration, \textit{Parton distributions for the LHC Run II}, \href{http://doi.org/10.1007/JHEP04(2015)040}{\textit{JHEP} \textbf{04} (2015) 040} [\href{https://arxiv.org/abs/1410.8849}{\texttt{arXiv:1410.8849}}] [\href{https://inspirehep.net/search?p=find+eprint+1410.8849}{{\tiny IN}SPIRE}].

\bibitem{FEWZ} R.~Gavin, Y.~Li, F.~Petriello and S.~Quackenbush, \textit{FEWZ 2.0: A code for hadronic Z production at next-to-next-to-leading order}, \href{http://doi.org/10.1016/j.cpc.2011.06.008}{\textit{Comput. Phys. Commun.} \textbf{182} (2011) 2388} [\href{https://arxiv.org/abs/1011.3540}{\texttt{arXiv:1011.3540}}] [\href{https://inspirehep.net/search?p=find+eprint+1011.3540}{{\tiny IN}SPIRE}].

\bibitem{ATLAS} ATLAS Collaboration, \textit{The ATLAS experiment at the CERN Large Hadron Collider}, \href{http://doi.org/10.1088/1748-0221/3/08/S08003}{\textit{JINST} \textbf{3} (2008) S08003} [\href{http://inspirehep.net/record/796888/references?ln=it}{{\tiny IN}SPIRE}].

\bibitem{CMS} CMS Collaboration, \textit{The CMS experiment at the CERN LHC}, \href{http://doi.org/10.1088/1748-0221/3/08/S08004}{\textit{JINST} \textbf{3} (2008) S08004} [\href{http://inspirehep.net/record/796887?ln=en}{{\tiny IN}SPIRE}].

\bibitem{Calame} C.M.C.~Calame, M.~Chiesa, H.~Martinez, G.~Montagna, O.~Nicrosini, F.~Piccinini and A.~Vicini, \textit{Precision measurement of the W-boson mass: Theoretical contributions and uncertainties}, \href{http://doi.org/10.1103/PhysRevD.96.093005}{\textit{Phys. Rev. D} \textbf{96} (2017) 093005} [\href{https://arxiv.org/abs/1612.02841}{\texttt{arXiv:1612.02841}}] [\href{https://inspirehep.net/search?p=find+eprint+1612.02841}{{\tiny IN}SPIRE}].

\bibitem{Halley} J.P.~Boyd, \textit{Finding the zeros of a univariate equation: proxy rootfinders, Chebyshev interpolation, and the companion matrix}, \href{https://doi.org/10.1137/110838297}{\textit{SIAM Rev.} \textbf{55}, 2 (2013) 375}.

\bibitem{Alioli} S.~Alioli et al., \textit{Precision studies of observables in $pp\to W\to l\nu_l$ and $pp\to Z,\gamma\to l^+l^-$ at the LHC}, \href{http://doi.org/10.1140/epjc/s10052-017-4832-7}{\textit{Eur. Phys. J. C} {\bf 77} (2017) 208} [\href{https://arxiv.org/abs/1606.02330}{\texttt{arXiv:1606.02330}}] [\href{https://inspirehep.net/search?p=find+eprint+1606.02330}{{\tiny IN}SPIRE}].





\end{thebibliography}
\end{document}